 \documentclass[preprint2]{aastex}

\usepackage{psfig,natbib}
\usepackage{apjfonts}
\usepackage{graphics}
\usepackage{psbox}

\newcommand{\kms}	{km~s$^{-1}$}
\newcommand{\h}   	{$h^{-1}\,$~kpc}
\newcommand{\mpc}       {$h^{-1}\,$~Mpc}

\newcommand{\apg}  	{^{>}_{\sim}}
\newcommand{\apl}  	{^{<}_{\sim}}

\newcommand{\lya}	{Ly$\alpha$}

\newcommand{\flux}      {ergs cm$^{-2}$ s$^{-1}$ \AA$^{-1}$}

\shorttitle{Ly$\alpha$ absorption around nearby galaxies.}
\shortauthors{D. V. Bowen et al.}

\received{xxx}
\accepted{xxx}

\begin{document}

\title{Ly$\alpha$ absorption around nearby galaxies.}
\author{David V.~Bowen$^{1}$, 
Max Pettini$^2$, and J. Chris Blades$^3$}

\affil{ $\:$}

\affil{$^1$ Princeton University Observatory, Princeton, NJ 08544} 

\affil{$^2$Institute of Astronomy, The Observatories, Madingley Rd.,
Cambridge CB3 0EZ, UK}

\affil{$^3$Space Telescope Science Institute, 3700 San Martin Drive, 
Baltimore, MD~21218}

\begin{abstract}

We have used STIS aboard HST to search for \lya\ absorption lines in
the outer regions of eight nearby galaxies using background QSOs and
AGN as probes. \lya\ lines are detected within a few hundred ~\kms\ of
the systemic velocity of the galaxy in all cases. We conclude that a
background line of sight which passes within $26-200$~\h\ of a
foreground galaxy is likely to intercept low column density neutral
hydrogen with $\log N$(H~I)$\apg\: 13.0$. The ubiquity of detections
implies a covering factor of $\simeq 100$~\% for low $N$(H~I) gas
around galaxies within 200~\h . We find, however, that the \lya\ lines
are usually composed of individual components spread out in
velocity over ranges of $300-900$~\kms .  Two sightlines show
components which are unusually broad for low-redshift \lya\ systems,
with Doppler parameters $\sim 150$~\kms. These may arise in intragroup
gas at temperatures of $1-2\times10^{6}$~K.

We discuss the difficulty in trying to associate individual absorption
components with the selected galaxies and their neighbors, but show that by
degrading our STIS data to lower resolutions, we are able to reproduce
the anti-correlation of \lya\ equivalent width and impact parameter found
at higher redshift. The anti-correlation does not improve by correcting for
the absolute magnitude of a galaxy in the same way as found at higher-$z$. We
also show that the equivalent width and column density of \lya\ complexes
(when individual components are summed over $\sim 1000$~\kms ) correlate
well with a simple estimate of the volume density of galaxies brighter than
$M_B = -17.5$ at the same redshift as a \lya\ complex. We do not reject the
hypothesis that the selected galaxies are directly responsible for 
the observed \lya\ lines, but our analysis indicates that absorption by clumpy
intragroup gas is an equally likely explanation.

\end{abstract}

\keywords{quasars:absorption lines---galaxies:halos --- intergalactic
medium --- large-scale structure of universe}

                \section{Introduction}

The initial detection of $z<< 1$ \lya -forest absorption lines in HST
QSO spectra \citep{Bah91, morr91, KP1} not only demonstrated the
existence and evolution of diffuse H~I clouds over a significant
fraction of the age of the universe, but made it possible to search
for the origin of the clouds themselves. Although \lya\ absorbers were
generally accepted
to be intergalactic clouds at high-redshift \citep{sarg80}, the success at
detecting galaxies associated with higher H~I column
density Mg~II systems at $z < 1$
\citep{jb1,crist87,BergBois,SteiMg2,stei95conf} fuelled a keen interest in
investigating whether or not low column density \lya\ lines might also
arise in the regions of individual galaxies.

The mapping of galaxies around the sightline towards 3C~273 \citep{salz92,
Morr93, salp95} found little evidence for a direct association between
galaxy and \lya -absorber. \citet{Morr93} concluded that \lya\ clouds are
not distributed at random with respect to galaxies, nor do they cluster as
strongly as galaxies cluster with each other, and could only be associated
with galaxies on scales of $\sim 0.5-1\:
h^{-1}$~Mpc\footnote{$h\:=\:H_0/100$, where $H_0$ is the Hubble constant,
and $q_0\:=\:0$ is assumed throughout this paper}. However, in a study of
six different fields, \citet[][hereafter LBTW]{lbtw} found that {\it i)}
the majority of normal, luminous galaxies possess extended \lya -absorbing
halos or disks of radii $\sim 160$~\h , and {\it ii)} between one and two
thirds of all \lya\ lines arise in such galaxies. Combined with other
similar studies \citep{Stoc95,LeBr96} it appeared that one possible way to
reconcile these disparate results was if strong \lya\ lines (those
with equivalent widths $W\apg 0.3$~\AA ) were associated
with galaxies, while weak ones ($W\apl 0.3$~\AA ) arose in the
intergalactic medium. Such a claim was quickly rejected by other authors
\citep[e.g.][]{Trip98}.

These initial results were soon re-evaluated in light of the rapid
development in hydrodynamical and semi-analytic simulations of hierarchical
cold dark matter structure formation.  Simulations showed  gas
following the same density fluctuations that are gravitationally induced by
dark matter distributions, resulting in a `web' of intersecting
filaments and sheets of gas. Analyses of artificial spectra, generated by
shooting random sightlines through the simulations, were extremely
successful in reproducing the observed properties of the high-redshift \lya
-forest \citep{cen94,zhang95,hern96,mira96,bryan99}. In particular, they
showed that low column density lines can be produced predominantly in the
sheets of gas, while higher column density lines can arise from denser
gas in a virialized halo, i.e., the same regions in which a high galaxy
density might be expected.

More recently,
\citet[hereafter CLWB]{chen98} and
\citet{chen01} have extended the original observations of LBTW, and continue to
find that the strength of the absorption depends on QSO-galaxy
separation (`impact parameter') as well as galaxy luminosity, suggesting
a strong link between galaxy and absorber.  Meanwhile, the simulators have
advanced their models to $z\sim 0$, and are again able to reproduce
many features of the observed \lya -forest \citep{theuns98,dave+tripp}.
\citet{dave99} have identified clumps of gas
and stars in their simulations which are likely to correspond to galaxies,
and impressively, have been able to reproduce the anti-correlation of line
strength and impact parameter.

In a previous paper \citep{BBP96} we used Archival HST Faint Object
Spectrograph (FOS) data to search
for \lya\ lines from {\it present-day} galaxies.  Studying the
absorption properties of galaxies in the local universe is useful because a
galaxy's properties are more easily determined and because its
environment can be examined in detail, without the difficulties which
naturally arise from observing higher redshift galaxies.  In the previous
paper, we found that nearby galaxies do not possess \lya -absorbing halos
beyond 300~\h\ in radius, and --- assuming that an individual galaxy was
indeed responsible for the absorption --- that the covering factor of
galaxies between 50 and 300~\h\ was $\sim 40$~\% at an equivalent width
limit $\geq 0.3$~\AA . However, we found no anti-correlation of \lya\ equivalent
width with impact parameter or with galaxy luminosity, and questioned
whether the galaxies were indeed responsible for the absorption lines. We
instead concluded that our results supported the emerging picture that
\lya\ lines arise in the gaseous sheets and filaments discussed above.

Unfortunately, our analysis suffered from two deficiencies: we probed few galaxies within
the canonical 160~\h\ found originally by LBTW and confirmed by CLWB; and
we were restricted to looking only for strong lines in the low resolution
FOS data. In this paper we seek to remedy these inadequacies by describing
the results of a program designed specifically to search for {\it weak}
\lya\ lines {\it within} 160~\h\ of a nearby galaxy.  The experiment was
not originally designed to address the origin of {\it all} \lya\ absorbers,
since we start by identifying a suitable galaxy and then search for \lya\
absorption at the galaxy's systemic redshift.  As we shall see, however,
this simple attempt to define galaxy-absorber associations is not so
straightforward, since the high spectral resolution of our STIS
observations not only enables us to identify lines weaker than those found
in FOS data, but also to resolve multiple components within an absorption
complex on a scale of $\approx 15$~\kms. As we will discuss, resolving
individual components in \lya\ systems makes it surprisingly difficult to
tie absorption lines to individual galaxies.

The paper is ordered as follows: We describe the selection of the
probed galaxies in \S\ref{sect_select}, and the 
data reduction and analysis in \S\ref{sect_analysis}.
 \S\ref{sect_results} details the geography of individual galaxy fields
and describes the \lya\ absorption lines detected in the STIS spectra.
\S\ref{sect_corrs} attempts to interpret the data obtained as a whole, and
\S\ref{sect_conclusions} summarizes our results and places them in context
with the higher-redshift systems discussed above.

        \section{Selection Criteria for the Observed Sample \label{sect_select}}

\subsection{Derivation of sample }

In order to produce a sample of QSO-galaxy pairs which could be
observed with HST, we cross-correlated the Third Reference Catalogue
of Bright Galaxies \citep[RC3;][]{RC3} with version 7 of the QSO/AGN
catalog of \citet{veron7} to find galaxies within 160~\h\ of a QSO/AGN
line of sight.  Selecting from the RC3 has the advantage that galaxies
have well-determined systemic velocities, often measured from the
21~cm emission line, as well as magnitudes determined in a consistent
way for all the galaxies.  We chose galaxies with velocities $ >
1300$~\kms , since absorption at velocities less than this would
likely be blended with the damped \lya\ absorption profile from the
Milky Way. Our sample was also supplemented with known QSO-galaxy
pairs available in the literature.

Initially, we attempted to select galaxies which---from a visual inspection
of STScI Digitized Sky Survey (DSS) images---appeared to be isolated from
other galaxies. Our aim was to test if single galaxies were surrounded by
\lya\ absorbing halos and avoid complications that might arise from
selecting galaxies with neighbors that might also have extended halos.
Galaxies are rarely found alone, however, and in retrospect this criterion
has largely been violated (see \S\ref{sect_results}).  One exception, where
we knew from the outset that a galaxy was in a group, was the field
containing NGC~3613 and NGC~3619, which lie close to the sightline towards
MCG+10$-16-111$. This AGN was eventually included due to the rarity of
finding a bright background probe so close to a foreground galaxy.  Also,
the galaxy group around NGC~6654A, which is $\sim 200$~\h\ away from the
sightline towards Q1831+731, and thus slightly beyond the 160~\h\ limit
initially imposed, was still included in our final sample since the QSO was
already selected due to its proximity to NGC~6654.

Finally, we imposed several more practical constraints: we favored probes
within HST's Continuous Viewing Zone to double the length of the available
exposure times, and preference was given to QSO and AGN with known UV
fluxes in order to estimate accurate exposure times for the required
signal-to-noise (S/N) of the observations. As will be seen from the data,
uniform S/N was not obtained for all the spectra, since the far-UV flux was
known for only three QSOs.

The QSO-galaxy pair Mrk~1048 and NGC~988 was initially selected as part of
our sample. However, prior to the execution of our program, we found that
the AGN was already due to be observed by another group (GO program 7345),
and we therefore removed it from our observing plan.  However, since the
pair were initially selected, we include the data here, extracted from the
HST Archive.

\subsection{Physical parameters of sample galaxies}

The list of probes observed in our program is given in
Table~\ref{tab_qsos}.  The galaxies foreground to these QSOs and AGN are
listed in Table~\ref{tab_gals}.  Most parameters in
Table~\ref{tab_gals} are taken directly from the RC3, although G1341+2555
is not listed therein, and data are taken instead from \citet{bowen95}.
Galaxy morphology is taken from the NASA Extragalactic Database (NED).

Three probes turned out to have far-UV fluxes considerably less than
expected from their visual magnitudes. ISO~1475+3554 was selected due to
its proximity on the sky to the bright elliptical galaxy UGC~1308, 8.2
arcmins or 118~\h\ away. SBS~1127+574 was observed due to its small
separation from NGC~3683A, 12.6 arcmins or 88~\h\ from the AGN sightline.
The X-ray source MS~10473+35, lies 29.1 arcmin, or 137~\h\ from NGC~3381,
and was potentially a good probe of the face-on galaxy.
All three probes had fluxes close to zero, rendering their data
useless. They are listed in Table~\ref{tab_qsos}, but are not included in
any further discussion.

\begin{deluxetable}{llcccclcrc}
\tabletypesize{\scriptsize}
\tablecolumns{10} 
\tablewidth{0pc} 
\tablecaption{Journal of Observations\tablenotemark{a} \label{tab_qsos}}
\tablehead{
\colhead{}       &  \colhead{}      & \colhead{}    & \colhead{}    &
\colhead{} & \colhead{}  & \colhead{}  & \colhead{} & \colhead{Exposure} & \colhead{} \\ 
\colhead{}       &  \colhead{}      & \colhead{RA \& DEC}   &
\colhead{Galactic}  & \colhead{}    & \colhead{} & \colhead{ } &\colhead{Observation}  
& \colhead{Time}     & \colhead{Observed} \\
\colhead{Target} & \colhead{Alias}  & \colhead{(J2000.0)} & \colhead{$l$ \& $b$} & \colhead{$V$} & \colhead{$z_{\rm{em}}$} & \colhead{Type}   & \colhead{Date} 
& \colhead{(min)}    &  \colhead{Flux\tablenotemark{b}}
}
\startdata
ISO 1475+3554    &  ...             & 01:50:32.74 $+$36:09:25.8 & 136.1 $-$25.2     & 15.3 & 0.080   & Sy2   & 06-Sep-00 & 88.7   & $\leq 0.01$ \\
Mrk 1048         & NGC 985          & 02:34:37.72 $-$08:47:15.7 & 180.8 $-$59.5     & 14.3 & 0.043   & Sy1   & 17-Feb-99 & 62.2   & \phm{$\leq$}5.0    \\
PKS 1004+130     & 4C+13.41         & 10:07:26.09 $+$12:48:56.2 & 225.1 $+$49.1     & 15.2 & 0.240   & QSO   & 23-Mar-00 & 86.5   & \phm{$\leq$}0.9    \\
MS 10473+35      & CGCG 184$-$023   & 10:50:10.86 $+$35:02:01.6 & 188.4 $+$63.0     & 15.6 & 0.040   & Sy1.9 & 20-May-99 & 138.4  & $\leq 0.04$ \\
ESO 438$-$G009   & \nodata          & 11:10:48.01 $-$28:30:03.7 & 277.5 $+$29.4     & 14.2 & 0.025   & Sy1.5 & 22-Mar-00 & 184.8  & \phm{$\leq$}0.1    \\
MCG+10$-16-$111  & SBS 1116+583A    & 11:18:57.72 $+$58:03:23.9 & 144.2 $+$55.1     & 15.7 & 0.027   & EmLS  & 02-Nov-99 & 325.0  & \phm{$\leq$}1.0 \\
SBS 1127+575     & ...              & 11:30:03.48 $+$57:18:29.5 & 142.9 $+$56.6     & 16.0 & 0.036   & Sy2   & 06-Sep-99 & 325.0  & $\leq 0.02$   \\
PG 1149$-$110    & \nodata          & 11:52:03.45 $-$11:22:24.4 & 280.5 $+$48.9     & 15.5 & 0.049   & Sy1   & 22-May-00 & 134.5  & \phm{$\leq$}0.1    \\
PG 1341+258        & Ton 730          & 13:43:56.72 $+$25:38:47.5 & \phn 28.7 $+$78.2 & 16.6 & 0.087   & Sy1   & 28-Jul-99 & 135.7  & \phm{$\leq$}0.9    \\
Q1831+731        & IRAS F18315+7310 & 18:30:23.08 $+$73:13:10.0 & 104.0 $+$27.4     & 15.5 & 0.123   & Sy    & 20-Jul-00 & 97.4   & \phm{$\leq$}4.5    \\
\enddata 
\tablenotetext{a}{All data were taken using the G140M STIS
grating centered at 1222~\AA\ and the 52x0.2 aperture.}\\
\tablenotetext{b}{Flux at $\lambda=1230$~\AA\ in units of $1.0\times10^{-14}$~\flux }
\end{deluxetable}

\begin{deluxetable}{lllcrcrrcccccc}
\tabletypesize{\small}
\rotate
\tablecolumns{14}
\tablewidth{0pc} 
\tablecaption{Galaxies Probed by Background Targets \label{tab_gals}}
\tablehead{
\colhead{}         &  \colhead{}            & \colhead{} & \colhead{}  & \multicolumn{4}{c}{Separations}  
&  \colhead{}      &  \colhead{} &  \colhead{} &  \colhead{} \\
\cline{5-8}  
\colhead{}         & \colhead{Intervening}  & \colhead{} & \colhead{$v_{\rm{gal}}$}  & \colhead{$\rho$} 
& \colhead{$\rho$} & \colhead{$D_0$}        &  \colhead{}    & \colhead{}    & \colhead{} &  \colhead{}            & \colhead{$M_B$} &  \colhead{$S$} &  \colhead{$M_{\rm{H I}}$ } \\
\colhead{Target}   & \colhead{Galaxy}       & \colhead{Type} &  \colhead{(\kms )}
& \colhead{($'$)}   
& \colhead{(\h )}  & \colhead{($'$)}          & \colhead{2$\rho/D_0$} &
\colhead{$m_B$}  & \colhead{$\sigma(m_B)$}  & \colhead{$B_0^T$} &   \colhead{$-5\log h$}  &  \colhead{(Jy)} &  \colhead{($10^9 M_\odot$)} \\ 
\colhead{ (1)}   & \colhead{(2) }       & \colhead{(3)} &  \colhead{(4) }
&\colhead{(5) }       & \colhead{(6)}   
& \colhead{(7)}  & \colhead{(8)}          & \colhead{(9) } &
\colhead{(10)}    & \colhead{(11)}  &  \colhead{(12)} &  \colhead{(13)} & \colhead{(14)} 
}
\startdata
PKS 1004+130    & UGC 5454     & SABdm        & 2792           & 10.5  & \phn84  & 1.1   & 19.1  & 14.7 & 0.2 & 14.4    & $-17.9$ 
	& \phn4.4     & 0.8 \\
ESO 438$-$G009  & UGCA 226     & SB(s)m       & 1507           & 25.2  & 110     & 3.0   & 16.8  & 14.8 & 0.2 & 13.8    & $-17.1$ 
	& 26.7    & 1.4 \\
MCG+10$-16-$111 & NGC 3613     & E6           & 1987           & 4.5   & \phn26  & 3.2   & 2.8   & 11.8 & 0.1 & 11.7    & $-19.8$ 
	& ...    & ... \\
		& NGC 3619     & (R)SA(s)0+   & 1542           & 18.2  & \phn85  & 2.6   & 14.0  & 12.4 & 0.1 & 12.4    & $-18.5$ 
	& \phn5.5     & 0.3 \\
PG 1149$-$110   & NGC 3942     & SAB(rs)c pec & 3696           & 8.7   & \phn92  & 1.4   & 12.4  & 13.3 & ... & 13.8    & $-19.1$ 
	& \phn5.8     & 1.9 \\
Q1831+731       & NGC 6654     & (R')SB(s)0/a & 1821           & 27.2  & 143     & 2.7   & 20.1  & 12.6 & 0.2 & 12.4    & $-18.9$ 
	& $<$17.2\phn\phn & $<1.4$\phn\phn \\
		& NGC 6654A    & SB(s)d pec?  & 1558           & 44.3  & 199     & 2.8   & 31.6  & 13.5 & 0.2 & 12.5    & $-18.5$ 
	& 17.8    & 1.0 \\
Mrk 1048        & NGC 988      & SB(s)cd      & 1504           & 36.3  & 158     & 4.7   & 15.4  & 11.9 & 0.2 & 10.9    & $-20.0$ 
	& 34.2    & 1.8 \\
\enddata
\tablecomments{Explanation of table entries: (1) target ID; (2)
foreground galaxy probed; (3) morphological type, taken from the NED; (4)
galaxy heliocentric velocity, all from 21 cm measurements, except NGC~3613; 
(5) QSO-galaxy
separation in arcmins; (6) physical distance of sightline from galaxy center
in \h\ derived from the listed galaxy velocity; (7) isophotal
major diameter at a surface brightness level of 25 mag arcsec$^{-2}$,
corrected to a face-on inclination, from the RC3;
(8) ratio of impact parameter to optical radius;
(9) apparent $B$-band magnitude from RC3; (10) error in $m_B$ from RC3;
(11) apparent face-on $B$-band magnitude, from RC3; (12) absolute $B$-band
magnitude derived from $B_T^0$ ; (13) 21~cm flux derived from RC3 entry $m_{21}$ with 
$\log S$(Jy) = 6.97 $-0.4m_{21}$; $3\sigma$ limit for NGC~6654 taken from \cite{eskpogge91};  
(14) H~I mass derived from $M_{\rm{H\,I}} = 2.36\times 10^5 
D({\rm{Mpc}})^2 S {\rm (Jy)} M_\odot$. 
}
\end{deluxetable}

             \section{Observations \& Data Reduction \label{sect_analysis}}

The HST STIS observations are journaled in Table~\ref{tab_qsos}.  All
data were taken with the G140M grating centered at 1222~\AA\ and the
52x0.2 aperture.  In most cases the spectra used were those extracted
with standard pipeline calibration. For sources with a weak
signal, we used the IRAF routine APALL to optimally extract data
and improve the S/N of the spectra. Unless otherwise stated in
\S\ref{sect_results},
all data were resampled to the original dispersion of 0.05~\AA\
pix$^{-1}$.

\begin{figure*}
{\psfig{figure=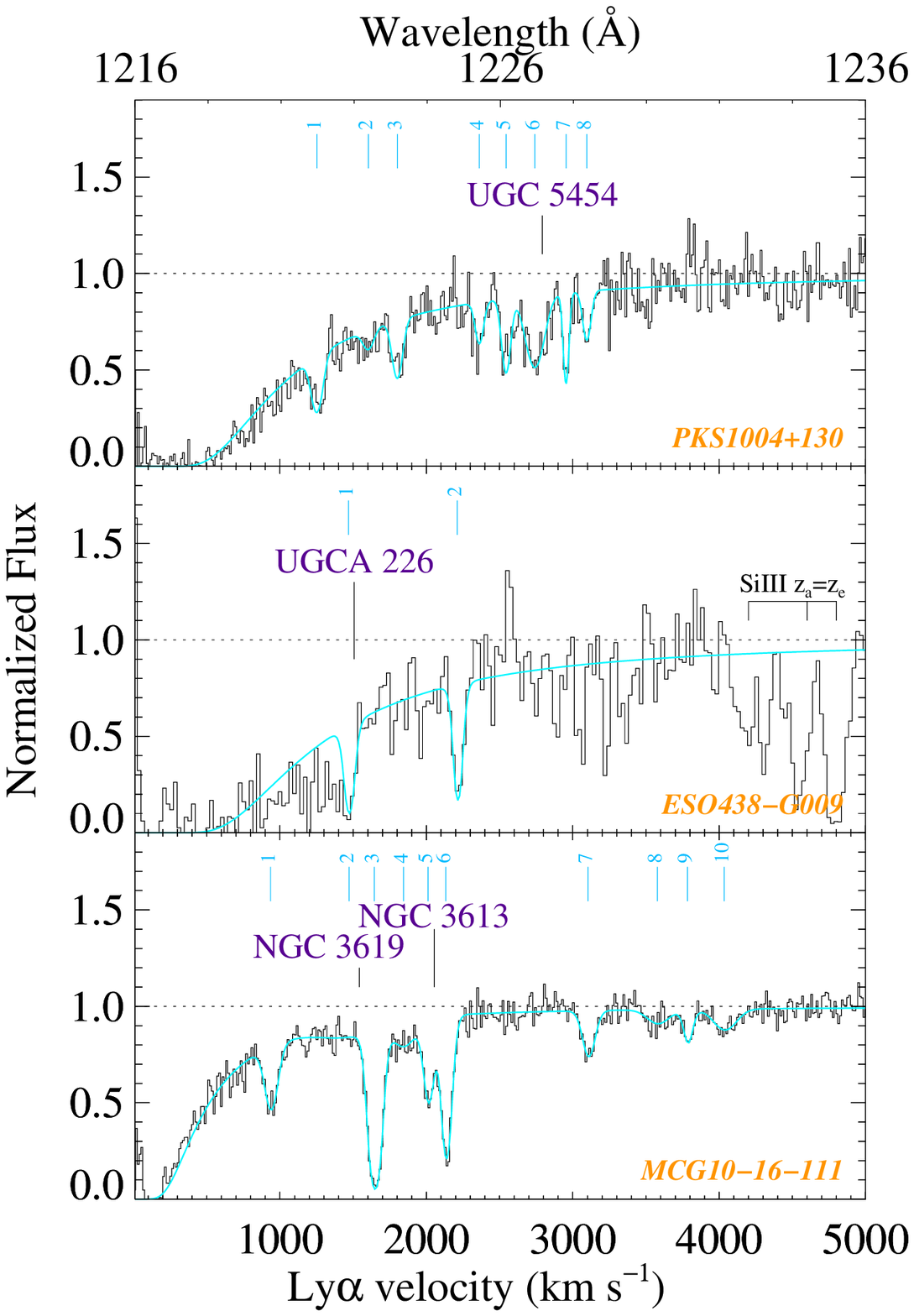,height=16cm,angle=0}}
\vspace*{-16.1cm}\hspace*{8.5cm}{\psfig{figure=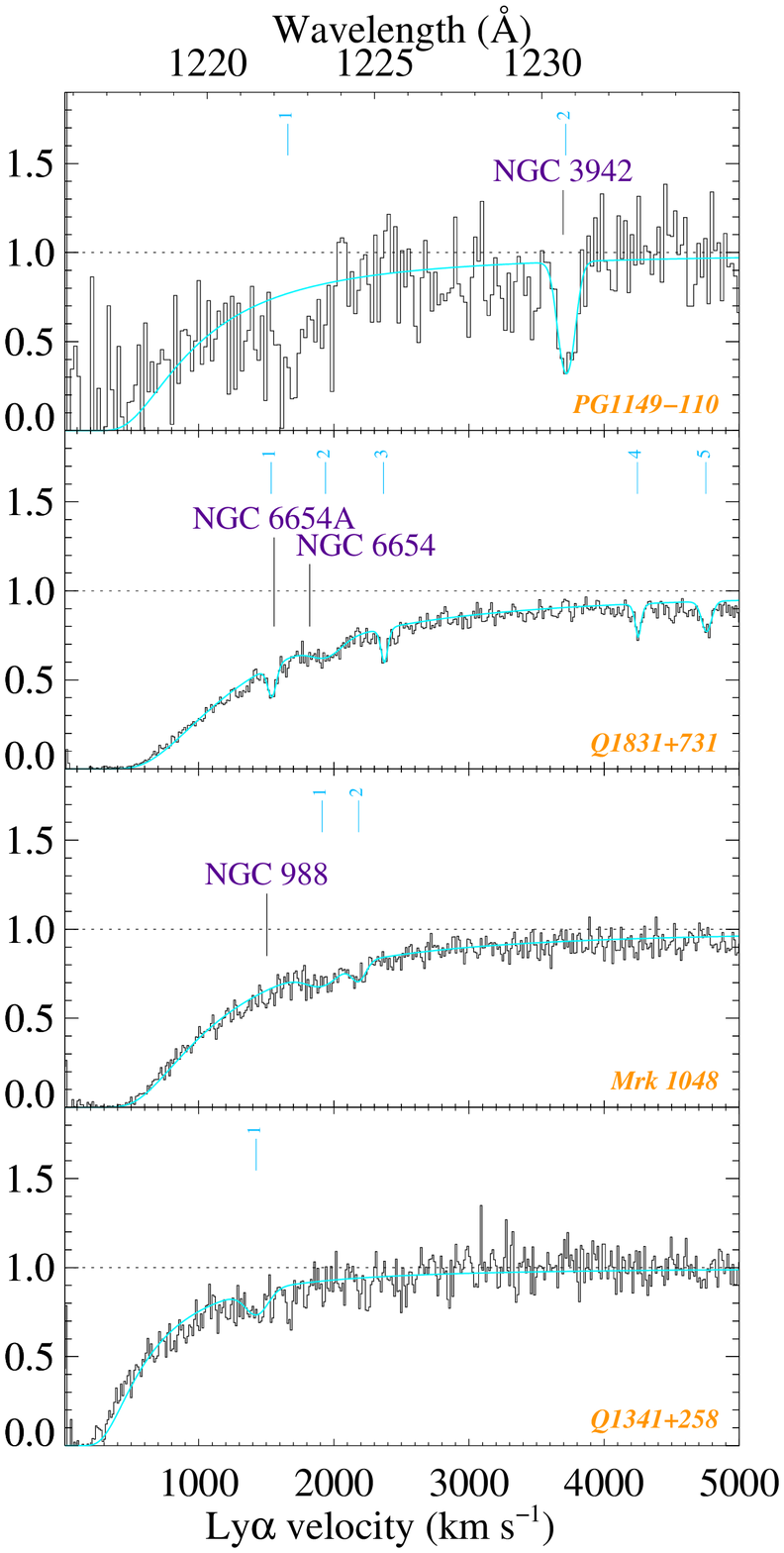,height=16cm,angle=0}}
\vspace*{-.5cm}\caption{\label{fig_spectra}{ STIS G140M spectra of the QSOs and AGN
reduced to a heliocentric velocity scale for \lya\ absorption
lines. 
The data are normalized, but the Milky Way damped \lya\ profile
stretching from zero to several thousand ~\kms\ has not been removed. For
ESO~438$-$G009, and PG~1149$-$110, the data have been
rebinned to 0.1~\AA~pix$^{-1}$, twice the original dispersion. The
velocities expected for \lya\ from individual galaxies are marked. 
Theoretical Voigt profiles representing our best fit to the lines are
overlaid, with their velocities indicated by tick-marks at
the top of each spectrum. No reliable fit could be produced for component 1 towards 
PG~1149$-$110. The data obtained towards PG 1341+258 are also shown, although the redshift of the galaxy 
originally selected to be close to the sightline was incorrect.}}
\end{figure*}

Normalization of the spectra was complicated by absorption from neutral
hydrogen in the Milky Way, which produces a damped \lya\ profile extending
over a significant velocity range. When the Galactic H~I column density,
$N$(H~I), is high, the QSO or AGN continuum may not return to its
unabsorbed value for several thousand \kms , which is a significant
fraction of the total wavelength covered by our observations with the G140M
grating.  To normalize the data in a way which retains the Galactic
absorption, a theoretical continuum must be fitted over a large wavelength
region where there is no information on the continuum's true value.  Hence,
to produce the normalized spectra in Figure~\ref{fig_spectra}, we required
that $N$(H~I) from the Milky Way absorption (as derived from our profile
fitting --- see below) match that derived from 21~cm emission
measurements in the Leiden/Dwingeloo H~I survey data
\citep{HIsurvey}. Although  H~I column densities measured from absorption
and emission line methods need not be the same (due, e.g., to beam smearing
in the 21~cm data), we found that the cores of the damped profile could be
reproduced well when adopting $N$(H~I) from the radio data.

Of course, analysis of any narrow extragalactic absorption lines need not rely on
data normalized this way, since it is always possible to normalize the
narrow lines themselves relative to whatever continuum can be placed either
side of a line. (That is, the data herein can be normalized in a way to
simply remove the Milky Way \lya\ absorption.)  However, some 
extragalactic \lya\ lines detected in our survey were at velocities
which placed them on the damping wings of the Milky Way \lya\ profile where
the change in flux with wavelength was rapid. We found that an `a priori'
knowledge of where the continuum should go --- as defined by the $N$(H~I)
column inferred from the 21~cm data --- was useful when the S/N of
the data was poor and the continuum either side of a weak line was
ill-defined. 

Measurements of the extragalactic lines themselves were made with the Galactic
H~I profile divided out of the data, although in Figure~\ref{fig_spectra} we
show the data with the Galactic absorption still present.  Equivalent
widths, $W$, of detected \lya\ lines were measured in the usual way \citep[see,
e.g.][]{bowen95} and are listed in Table~\ref{tab_cols}. 
We measure equivalent widths of individual components whenever possible,
but where lines are blended, we calculate $W$ over several components, indicating
which components are covered in the column headed `$\Sigma$'.

We also fitted
theoretical Voigt profiles to the data, using a Line Spread Function
(LSF) appropriate for the G140M grating at 1200~\AA\ and the 52x0.2
aperture. This LSF is composed of a sharp narrow Gaussian core with a FWHM
$\sim 14$~\kms , convolved with a wide, shallow profile which can be
represented by a Gaussian with a FWHM of $\simeq 64$~\kms .  We fitted all
\lya\ lines detected, not just those at velocities near the systemic
velocities of the galaxies of interest, up to $\sim 5000$~\kms. 
Resulting recession velocities, $v$, H~I column densities, $N$(H~I), and
Doppler parameters, $b$, are listed in Table~\ref{tab_cols}, along with
the resulting theoretical equivalent widths, $W_i$, derived from the final
values of $b$ and $N$(H~I).

Errors in these values, $\sigma(v)$, $\sigma(N)$, $\sigma(b)$ are also
given in Table~\ref{tab_cols}. These were calculated using a Monte Carlo
type simulation, described in detail in an earlier paper \citep{bowen95}.
To recap, we take the synthetic spectrum derived from the $\chi^2$ best fit
to the data as a template, and from this generate 100 new spectra with the
same S/N as the original data. Each of these spectra are then refitted with
theoretical Voigt profiles, again minimizing $\chi^2$. Each spectrum
generates a new value of $v$, $b$, and $N$(H~I). The distribution of $v$
values is usually Gaussian, so the values of $\sigma$ listed in
Table~\ref{tab_cols} are those for a normal error distribution. Errors
in $b$ and $N$ need not be independent of each other, however. For example,
a narrow line which covers only a few pixels is more sensitive to sudden
changes at the bottom of the line profile, resulting in a dramatic increase
in $N$(H~I) and a decrease in $b$. The exact way in which $N$(H~I) and $b$
change in relation to each other for different spectra depends on the
initial strength of a line, the S/N of the data, and the degree to which
a line is resolved. Where the distribution of $N$(H~I) and $b$ are
clearly not Gaussian, we simply take $1\sigma$ to be the value of $\Delta
N$(H~I) and $\Delta b$ for which 68\% of the points lie away from the mean
value of $N$(H~I) and $b$. Where $N$(H~I) and $b$ are strongly correlated,
$+\sigma$ deviations in one quantity are always associated with $-\sigma$
deviations in the other quantity.

Obviously, these errors only take into account uncertainties arising
from Poisson statistics, since the template spectrum itself
suffers from the same systematic errors found in the data. Hence the
errors do not account for incomplete knowledge of the multicomponent
structure of a cloud, an erroneous LSF, or (for the derived values
of $v$) errors in the wavelength calibration of STIS data. This
latter quantity should be small; according to the STIS Handbook, the
error in the absolute wavelength of spectra should be no more than
$0.2-0.5$ pixels, or 2$-$6~\kms . Errors arising from a poorly
determined LSF will be small providing the line is resolved. When this
is not the case, errors in $N$ and $b$ are large anyway, 
since $N$ and $b$ cannot be constrained with
only one unresolved \lya\ line. (Such uncertainties are accounted for
in the monte-carlo simulations, since $N$ and $b$ can deviate widely
from their initial values with no discernible variation in a line
profile.)

\begin{deluxetable}{llcccccccccrcrc}
\tabletypesize{\small}
\rotate
\tablecolumns{15}
\tablewidth{0pc} 
\tablecaption{\lya\ absorption lines detected in G140M STIS spectra \label{tab_cols}}
\tablehead{
\colhead{}       & \colhead{}             &  \colhead{}             &  \colhead{}            & \colhead{}    & \colhead{}     
& \colhead{} & \multicolumn{8}{c}{Profile fitting $N$(H~I)} \\
\cline{7-15}
\colhead{}       & \colhead{Intervening}  & \colhead{$v_{\rm{gal}}$} & \colhead{$W$}   & \colhead{$\sigma(W)$}  
& \colhead{}     & \colhead{}     
& \colhead{ }    & \colhead{$v$}          & \colhead{$\sigma(v)$}    & \colhead{$b$}   & \colhead{$\sigma(b)$}  
& \colhead{}     & \colhead{ } & \colhead{ $W_i$\tablenotemark{a}} \\
\colhead{Target} & \colhead{Galaxy}       & \colhead{(\kms)}         & \colhead{(\AA )}        & \colhead{(\AA )}       
& \colhead{$\Sigma$\tablenotemark{b}} & \colhead{}    
& \colhead{\#}     & \colhead{(\kms)}       & \colhead{(\kms)}         & \colhead{(\kms)}        & \colhead{(\kms)}         
& \colhead{$\log N$(H~I)} & \colhead{$\sigma  (\log N)$} & \colhead{(\AA )}
}
\startdata
PKS 1004+130    & UGC 5454  & 2792  & 0.22 & 0.02 & 1   &&1 & 1252 & 5  & 49  & 10           & 13.67 & 0.05 &0.20 \\
                &           &       & 0.08 & 0.03 & 2   &&2 & 1604 & 7  & 51  & 15           & 13.02 & 0.06 & 0.05 \\
                &           &       & 0.13 & 0.02 & 3   &&3 & 1802 & 7  & 46  & 15           & 13.49 & 0.09 & 0.14 \\
                &           &       & 0.67 & 0.05 & 4-8 &&4 & 2362 & 5  & 37  & 13           & 13.16 & 0.20 & 0.07 \\
                &           &       &      &      &     &&5 & 2545 & 7  & 38  & 13           & 13.48 & 0.08 & 0.14 \\
                &           &       &      &      &     &&6 & 2741 & 5  & 83  & 13 	     & 13.76 & 0.05 & 0.26 \\
                &           &       &      &      &     &&7 & 2954 & 3  & 22  & 15           & 13.38 & 0.20 & 0.10 \\
                &           &       &      &      &     &&8 & 3096 & 8  & 36  & 16 	     & 13.20 &$^{+0.20}_{-0.15}$ & 0.08\\
ESO 438$-$G009  & UGCA 226  & 1507  & 0.51 & 0.20 & 1   &&1 & 1469 & 5  & 30: & 10 	     & 14.13:& 0.2: & 0.31 \\
                &           &       & 0.23 & 0.04 & 2   &&2 & 2211 & 6  & 25  &$^{+13}_{-10}$& 14.08 &$^{+2.38}_{-0.73}$ & 0.25 \\
MCG+10$-16-$111	& NGC 3619  & 1542  & 0.20 & 0.01 & 1   &&1 & 936  & 3  & 51  & 4  	     & 13.64 & 0.02 & 0.19 \\
		&           &       & 0.62 & 0.02 & 2-4 &&2 & 1472 & 84 & $\geq300$& 50      & 13.48 & 0.12 & 0.10  \\
		&           &       &      &      &     &&3 & 1645 & 1  & 40  & 2            & 14.45 & 0.04 & 0.47 \\
                & NGC 3613  & 1987  & 0.54 & 0.02 & 4-6 &&4 & 1844 & 24 & 101 &$^{+79}_{-35}$& 13.33 & 0.14 & 0.11 \\
		&           &       &      &      &     &&5 & 2012 & 3  & 38  & 6            & 13.61 & 0.05 & 0.17 \\
		&           &       &      &      &     &&6 & 2133 & 1  & 32  & 3            & 14.04 & 0.03 & 0.29 \\
		&           &       & 0.10 & 0.02 & 7   &&7 & 3104 & 6  & 49  & 11           & 13.32 & 0.05 & 0.10 \\
		&           &       & 0.11 & 0.02 & 8-9 &&8 & 3578 & 30 & 98  &$^{+39}_{-47}$& 13.00 &$^{+0.17}_{-0.13}$ & 0.05 \\
		&           &       &      &      &     &&9 & 3783 & 7  & 31  & 14           & 13.04 & 0.09 & 0.05 \\
		&           &       & 0.09 & 0.01 & 10  &&10& 4034 & 18 & 103 & 29           & 13.20 & 0.08 & 0.08 \\
PG 1149$-$110   & NGC 3942  & 3696  & 1.1  & 0.3  & 1   &&1 & 1660 & 15 & ... & ...	   & ...   & ...  & ... \\
		&           &       & 0.39 & 0.10 & 2   &&2 & 3716 & 14 & 71  &$^{+44}_{-25}$& 14.04 &$^{+0.18}_{-0.12}$ & 0.41 \\
PG 1341+258       & ...       & ...   & 0.12 & 0.02 & 1   &&1 & 1425 & 16 & 110 & 18           & 13.40 & 0.08  & 0.13 \\
Q1831+731       & NGC 6654A & 1558  & 0.11 & 0.02 & 1   &&1 & 1536 & 3  & 32  & 5  	      & 13.33 & 0.03 & 0.10 \\
                & NGC 6654  & 1821  & 0.11 & 0.02 & 2   &&2 & 1938 & 13 & 148 & 20           & 13.38 & 0.05 & 0.12 \\
                &           &       & 0.10 & 0.02 & 3   &&3 & 2368 & 4  & 21  & 6            & 13.15 & 0.05 & 0.06 \\
                &           &       & 0.05 & 0.01 & 4   &&4 & 4246 & 4  & 15  & 10           & 13.00 &$^{+0.12}_{-0.08}$ & 0.05 \\
                &           &       & 0.09 & 0.05 & 5   &&5 & 4752 & 5  & 40  & 8            & $\geq13.11$ & 0.05 & 0.07 \\
Mrk 1048        & NGC 988   & 1504  & 0.12 & 0.01 & 1   &&1 & 1913 & 23 & 162 & 33           & 13.43 & 0.09 & 0.13 \\
                &           &       & 0.08 & 0.01 & 2   &&2 & 2183 & 17 & 67  & 27           & 13.12 & 0.13 & 0.07 \\
\enddata
\tablenotetext{a}{The equivalent width of a component derived from the values of $b$ and $N$}\\
\tablenotetext{b}{The components (from column 7) included in the wavelength range over which the equivalent width was calculated.}
\end{deluxetable}

Most lines measured in our data appear to be resolved, suggesting that
$N$(H~I) is reasonably well constrained.  Of course, we cannot exclude the
possibility that lines are actually comprised of several
components, with broad lower column density absorption lines masking narrow higher
$N$(H~I) components.

		\section{Results \label{sect_results}}

In the following subsections we outline the available data on each
QSO-galaxy field studied.

Although we selected sightlines to pass
close to nearby galaxies, we also have the opportunity to define the
environment of these galaxies because of their low redshifts.
Indeed, given the interest in whether or not \lya\ clouds trace the same 
large scale structure as galaxies, it seems incumbent on us to  explore any
relationship which might exist between such structures and \lya\
absorbers. Unfortunately, our experiment was not initially designed to encompass such
inquiries, and 
to proceed further we need to use published galaxy catalogs.
To this end, we have selected all galaxies from the
electronic version of the
RC3\footnote{Catalog 7155, as distributed by the Astronomical Data Center
(ADC) at NASA Goddard Space Flight Center.} which lie within 691 arcmins of
each QSO/AGN. This radius corresponds to 2~\mpc\ at a redshift of 1000~\kms
, roughly the cut-off in velocity for which \lya\ can be seen before being
lost in the damped \lya\ profile of the Milky Way.  The RC3 is ``reasonably
complete for galaxies having apparent diameters larger than 1 arcmin at the
$D_{25}$ isophotal level and total $B$-band magnitudes $B_T$ brighter than
about 15.5'' \citep{RC3}.  We have also searched the NED within a radius of
300 arcmin (the maximum search radius allowed) or 868~\h\ at 1000~\kms ,
for all galaxies with known redshifts, irrespective of their magnitudes.

Obviously, neighboring galaxies collated in this way do not form a
complete magnitude limited survey. For example, a lack of redshifts in
any given field does not necessarily imply that no galaxies exist,
merely that no measurements of their redshifts have been
made. Similarly, any peak in the galaxy distribution at low redshift
may be misleading, since the absence of galaxies beyond a few thousand
\kms\ may again merely reflect the fact that a particular area of sky
has not been surveyed to any depth in redshift.  It is therefore
important to realize that the histograms shown in \S\ref{sect_results}
represent a `first-look' at the distribution in redshift of all
galaxies with known redshifts. We defer a more rigorous analysis of
these distributions until \S\ref{sect_corrs}.

\subsection{PKS1004+130 \& UGC 5454 \label{sect_pks1004}}

\begin{figure*}
\vspace*{-2cm}\centerline{\psfig
{figure=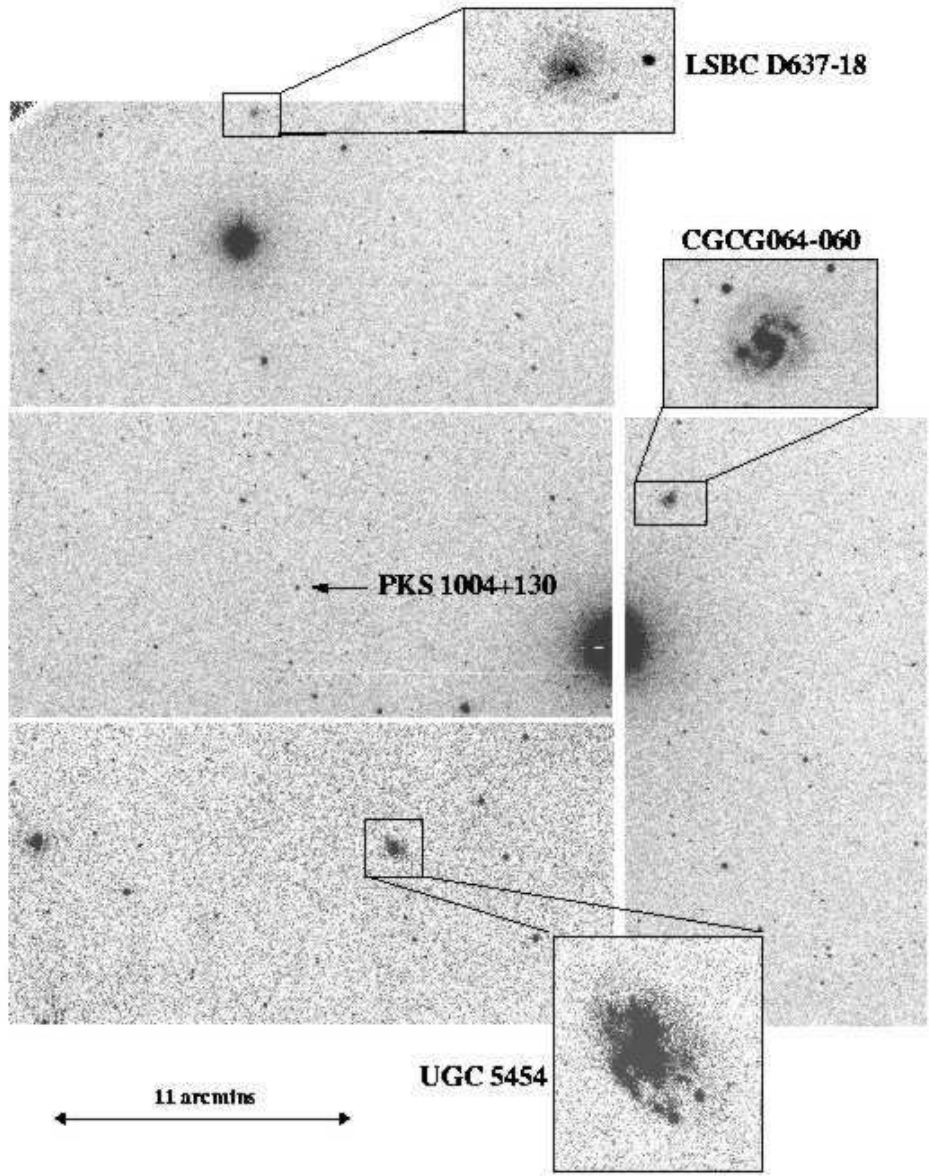,height=16cm,angle=0}}
\caption{\label{fig_PKS1004wfc} A {\it Wide Field Camera} image of the
field around PKS~1004+130 taken at the Isaac Newton Telescope on La
Palma. UGC~5454 ($M_B = -17.9$) lies 84~\h\
away from the QSO sightline. The low surface brightness galaxy LSBC D637$-$18
($M_B = -15.3$) lies 138~\h\ away. These two galaxies, along with a third
not covered in this image, CGCG~064$-$055
($M_B=-16.9$), which lies 198~\h\ away from the line of sight, appear
to constitute a group.
(CGCG~064$-$060 is a background galaxy
whose redshift is too high to permit us to search for \lya\ absorption.)
We detect  a complex \lya\ line at
the group's velocity, comprised of at least five components spread over
740~\kms .}
\end{figure*}

The sightline to PKS~1004+130 passes 10.5 arcmins or 84~\h\ from
UGC~5454, a SABdm galaxy with a velocity of $v=2792$~\kms .
Figure~\ref{fig_PKS1004wfc} shows a 200 sec $B$-band Wide Field Camera
(WFC) image of the field taken at the Isaac Newton Telescope on
24-Nov-1998.  The galaxy is irregular, with disturbed spiral arms
containing discrete H~II regions.

Two other galaxies are also visible in the CCD image.  CGCG064$-$060 lies
14.4 arcmins from the sightline, at a velocity of $9194$~\kms , which gives
a separation of 368~\h . The wavelength at which \lya\ might be expected
from this galaxy is not covered by our G140M observations.  The second
galaxy of interest is a low surface brightness galaxy designated by
\cite{pildis97} as D637$-$18\footnote{see also
http://zebu.uoregon.edu/$\sim$js/dwarfs/d637-18.ps for full details.},
which lies 17.6 arcmin from the QSO sightline, or 138.1~\h\ given its
velocity of $2740$~\kms\ (a redshift similar to that of UGC~5454).  There
also exists a third galaxy not shown in Figure~\ref{fig_PKS1004wfc} with
a velocity close to UGC~5454: CGCG~064$-$055 has $v=2789$~\kms\ and $\rho =
24.7$ arcmins $\equiv 198$~\h . We summarize the properties of all
additional galaxies within 200~\h\ of a sightline in
Table~\ref{tab_others}.

\begin{figure}[th]
\hspace*{-1.0cm}
\psbox[xsize=0.6\textwidth,rotate=l]{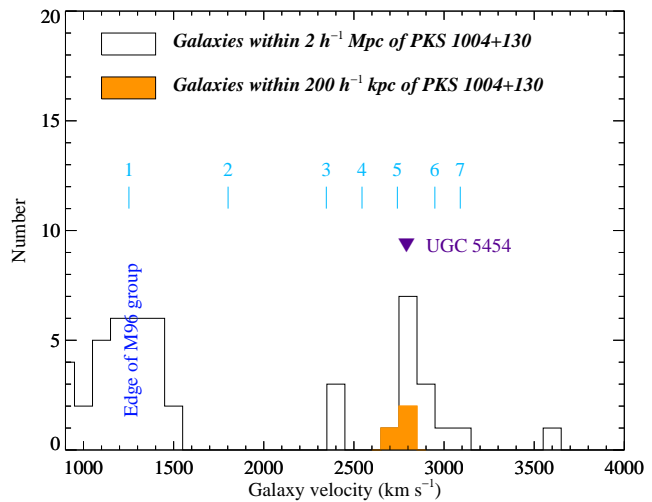}
\caption{\footnotesize \label{fig_pks1004dist}
Distribution of all galaxies with known redshifts within a distance of
2 \& 0.2~\mpc\ of the sightline to PKS~1004+130.  The velocity of our
selected galaxy UGC~5454 is marked with a $\blacktriangledown$.  The
complex of \lya\ absorption lines (components 3---7) appear to be
coincident with an over-density of galaxies out to 2~\mpc .}
\end{figure}

Collating galaxies with known redshifts reveals a strong over-density of
galaxies at a velocity similar to the UGC~5454
group. Figure~\ref{fig_pks1004dist} shows a histogram of galaxy velocities
for objects within 2~$h^{-1}$~Mpc of the line of sight.  A clear peak can
be seen at the same velocity as the three galaxies discussed above, making
it likely that the group around UGC~5454 is part of a larger conglomeration
of galaxies.

The nearest cataloged cluster of galaxies to the sightline is the M~96
group, which is 10 degrees away. The group has a velocity of 969~\kms ,
which gives a linear separation between QSO sightline and the group's
center of 1.7~\mpc. [Assuming a distance of 11.2~Mpc to M96
measured from Cepheids \citep{tanv99} gives a separation of 2.0~Mpc.] The
giant intergalactic cloud discovered by \cite{sch83} is also only 9.9
degrees away; the ring has a heliocentric velocity of 870$-$1050~\kms ,
with the bulk of the absorption at 960~\kms .  

The STIS spectrum shows a cluster of \lya\ absorption lines centered
on the systemic velocity of UGC~5454, with at least five components
spread over 740~\kms\ (see Table~\ref{tab_cols} for details).
\citet{wills99} have shown that there exists strong O~VI, N~V and
\lya\ at redshifts of $z=0.2364$ and $z=0.2387$ which probably arise
from outflows from the QSO. It is likely that we detect C~III~$\lambda
977$ at 1210.2~\AA\ from the $z=0.2387$ system, which gives rise to
the possibility that the \lya\ complex centered on the redshift of
UGC~5454 is contaminated by another strong, high ionization line,
N~III~$\lambda 989$ at 1226.1~\AA\ from this system. We note, however,
that if this line does exist, it will only contaminate (or perhaps be
responsible for) component 4 of the complex.

The velocity of the M~96 group and the giant H~I cloud is considerably
lower than that of UGC~5454 and its neighbors, but a \lya\ line
(component 1 in Figure~\ref{fig_spectra} and Table~\ref{tab_cols}) is
detected at 1252~\kms . This \lya\ cloud may be associated with the
outer regions of the M~96 group, or with some other member galaxy
close to the QSO sightline whose redshift remains undetermined.

There are three possible ways to understand the absorption detected.  As we
discuss in \S\ref{sect_fos}, at higher redshift the absorption towards
PKS~1004+130 would appear as a single strong resolved \lya\ line in the
lower resolution FOS data. In a search for a galaxy responsible for the
absorption, UGC~5454 would be the nearest to the sightline, but whether its
redshift would be measured is unclear, since it would be fainter at higher
$z$.  (For example, with an absolute magnitude of $M_B=-17.9$, the galaxy's
$R$-band magnitude would range from 19.6 to 22.2 at $z=0.1$ and
$z=0.3$. The other two nearby galaxies, with magnitudes $1-2$ times
fainter, would be even less likely to be identified at higher redshift.)
Assuming it was identified, the results from the surveys of CLWB and
\citet{chen01} would predict that UGC~5454 would be associated with a \lya\
line of strength $\geq 0.3$~\AA\ at a radii of 101~\h\ [using their value
of $M_B^* = -19.5$ and the relationship $\rho = 174 (L_B/L_B^*)^{0.37}$ \h
]. Our observations are consistent with this expectation: the integrated
\lya\ equivalent width is 0.68$\pm0.05$~\AA\ at an observed impact
parameter of 84~\h .

The higher-resolution STIS observations, however, show us that the
absorption line is actually a complex of individual components.
\cite{Orti99} have suggested that the complex system of \lya\ lines
towards Q1545+2101 observed with the GHRS aboard HST might arise in
the overlapping halos of individual galaxies in dense regions (in this
case, the cluster surrounding the observed QSO, since
$z_{\rm{abs}}\simeq z_{\rm{em}}$). With the field around
PKS~1004+130 well sampled (at least to the limit of the RC3), we have
a fairly good inventory of galaxies close to the line of
sight. The absorption systems at $z\simeq 0.265$ towards Q1545+2101
covers a range of 535~\kms , a little smaller than the 740~\kms\ seen
towards PKS~1004+130, but the complexity and velocity span 
suggests a similarity between the systems.

For PKS~1004+130, however, it seems harder to make the case that
overlapping halos of galaxies in the group generate the multi-component,
high-velocity spread within the \lya\ absorption system.  The galaxies in
the group are faint and distant from the sightline. For the absolute
magnitudes given in Table~\ref{tab_others}, the three galaxies would be
predicted by CLWB to have \lya -absorbing radii of $\sim$ 100, 42 and
72~\h\ . The latter two radii are smaller than the actual impact parameters
of the galaxies, i.e. the galaxies are too far away to produce the observed
lines. There are also at least five lines comprising the system, whereas we
only know of three galaxies within 200~\h\ of the sightline. On the other
hand, the predicted radii of galaxies at higher-$z$ assume that \lya\
lines would have strengths $\geq 0.3$~\AA .  Individual components
marked in Figure~\ref{fig_spectra} have equivalents widths ranging from
$0.07\pm0.02$~\AA\ (component 3) to $0.26\pm0.02$~\AA\ (component 5), all
less than 0.3~\AA . If the anti-correlation of $W$ and $\rho$ continues for
faint galaxies over large distances, then it is possible that more
distant galaxies contribute to the absorption complex. Nevertheless, the
fact that the group of three is not particularly rich, and that one galaxy
would have to produce more than one component to contribute to the five
observed, makes it less likely that the halos of individual galaxies in
the group produce the observed absorption.

There are two other possible explanations for the observed \lya\ complex.
Absorption might arise from a turbulent outflowing wind from UGC~5454. The
galaxy appears irregular and may be actively forming stars.  \citet{heck00}
have suggested that superwinds from active star-forming galaxies are
blowing metals into the intergalactic medium (IGM), enriching gas to a
metallicity as high as $\sim\, 1/6$ of the solar value.  However, the {\it
mean} velocity of the absorption lines towards PKS~1004+130 is not
displaced with respect to the galaxy's systemic velocity, as might be
expected from material flowing towards or away from us. The wind would
likely have to be expelled perpendicular to the plane of the sky to cause
absorption at both positive and negative velocities with respect to
UGC~5454's systemic velocity, as is observed.  Such an alignment is not
impossible of course, and further study of UGC~5454 is needed to determine
the nature of this galaxy.

Finally, Figure~\ref{fig_pks1004dist} shows that the group galaxies
discussed mark the presence of a strong over density of galaxies at
precisely the velocity of the center of the observed absorption line
complex. Since the large velocity dispersion is more in keeping with
that of a cluster, it is possible that the absorption arises from
intragroup or intracluster gas, material not necessarily located in
the same physical space as UGC~5454. The existence of faint galaxies
near the line of sight would then simply be a coincidence.

\begin{deluxetable}{lccccc}
\tablecolumns{6}
\tablewidth{0pc} 
\tablecaption{Galaxies with known redshifts within 200~\h\ of sightlines\tablenotemark{a} \label{tab_others}}
\tablehead{
\colhead{Intervening}  & \colhead{$v_{\rm{gal}}$}  & \colhead{$\rho$} 
& \colhead{$\rho$}     & \colhead{}                & \colhead{} \\
\colhead{Galaxy}       & \colhead{(\kms )}         & \colhead{($'$)}   
& \colhead{(\h )}      & \colhead{$m_B$}           & \colhead{$M_B-5\log h$}  
}
\startdata
\multicolumn{6}{c}{sightline: PKS 1004+130}\\
\hline
UGC 5454        & 2792            & 10.5  & 84.3    & 14.4    & $-17.9$ \\
LSBC D637$-$18  & 2740            & 17.6  & 138.1   &  ...    & $-15.3$        \\
CGCG 064$-$055  & 2789            & 24.7  & 197.6   & 15.3    & $-16.9$        \\ 
\hline
\multicolumn{6}{c}{sightline: ESO 438-G009}\\
\hline
 UGCA 226       &   1507      &   25.2    & 110        &  13.8   &  $-17.1$    \\
ESO 438$-$G012  &   1322      &   35.9    & 137        &  13.7   &  $-16.9$                \\
ESO 438$-$G010  &   1487      &   36.3    & 155        &  14.3   &  $-16.6$              \\ 
\hline
\multicolumn{6}{c}{sightline: MCG+10$-16-111$}\\
\hline
CGCG 291$-$052  & 9753            & 1.3   & 35.2    &   15.1    &  $-19.9$       \\ 
SBS 1116+583B   & 9905            & 3.6   & 98.8    &   19.5    &  $-15.5$      \\ 
NGC 3613        & 1987            & 4.5   & 25.5    &   11.7    &  $-19.8$       \\ 
NGC 3619        & 1542            & 18.2  & 81.0    &   12.4    &  $-18.5$       \\
UGC 6304        & 1762            & 19.8  & 100.6   &   16.5    &  $-14.7$       \\ 
NGC 3625        & 1940            & 20.7  & 115.7   &   13.2    &  $-18.2$       \\
UGC 6344        & 1934            & 22.1  & 122.9   &   17.     &  $-14.4$       \\ 
SBS 1119+583    & 1623            & 27.6  & 129.2   &   17.     &  $-14.1$       \\ 
SBS 1114+587    & 1594            & 31.3  & 144.0   &   17.5    &  $-13.5$       \\ 
SBS 1119+586    & 1583            & 33.3  & 152.1   &   19.5    & $ -11.5$       \\ 
\hline
\multicolumn{6}{c}{sightline: Q1831+731}\\
\hline
NGC 6654        & 1821            & 27.2  &  143   &  12.5       &  $-18.8$      \\
MCG+12-17-027   & 1404            & 43.7  &  177   &  17.0       &   $-13.7$      \\ 
UGC 11331       & 1554            & 43.7  &  196   &  15.4       &   $-15.6$       \\ 
NGC 6654A       & 1558            & 44.3  &  199   &  12.5       &  $ -18.5 $     \\
\hline
\multicolumn{6}{c}{sightline: Mrk~1048}\\
\hline
NGC 988         & 1504            & 36.3  & 158     & 10.9    &   $-20.0$      \\
\enddata
\tablenotetext{a}{Mrk~1048 and PG~1149$-$110 have no galaxies with known redshifts 
within 200~\h\  other than the objects already identified in Table~\ref{tab_gals}. These sightlines are
therefore not listed.}
\end{deluxetable}

\subsection{ESO 438-G009 \& UGCA 226}

\begin{figure}
\centerline{\psfig
{figure=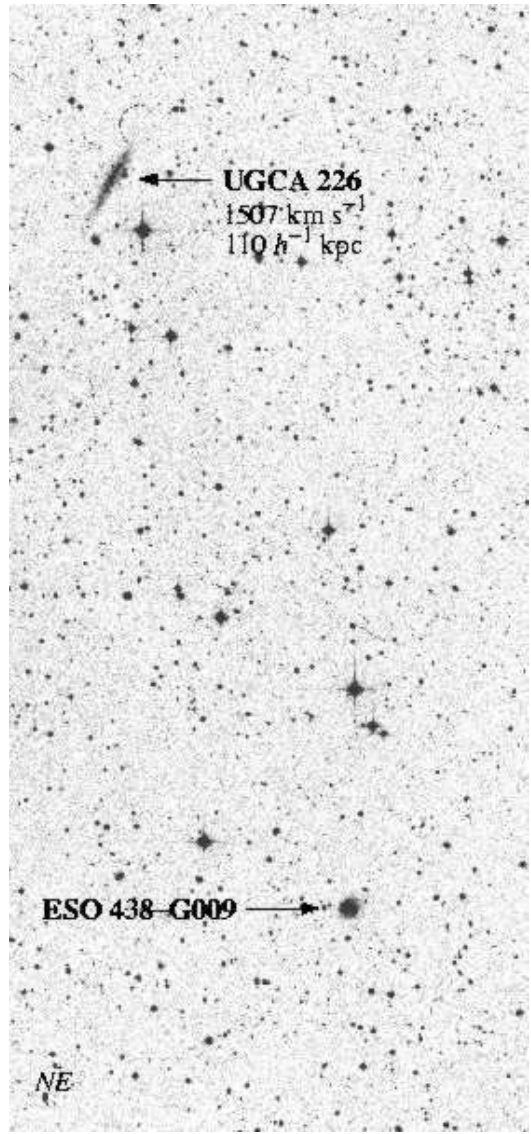,height=15cm,angle=0}}
\caption{\label{fig_ES438dss}\footnotesize STScI Digitized Sky Survey 
image of the field around
ESO~438$-$G009. 
UGCA~226 ($M_B=-17.1$) lies 110~\h\ from the line of
sight. Not shown are two other galaxies with similar velocities and small
distances from the QSO sightline: ESO 438$-$G012 ($v=1322$~\kms , $M_B =
-16.9$) is 137~\h\ away, while ESO 438$-$G010 ($v=1487$~\kms , $M_B = -16.6$) is
156~\h\ away. It may be that these three galaxies form a group. For scale,
the separation between ESO~438$-$G009 and UGC~226 is 25.2 arcmins.
\lya\ absorption is detected at a velocity of 1469~\kms . }
\end{figure}

\begin{figure}[th]
\hspace*{-1.0cm}
\psbox[xsize=0.6\textwidth,rotate=l]{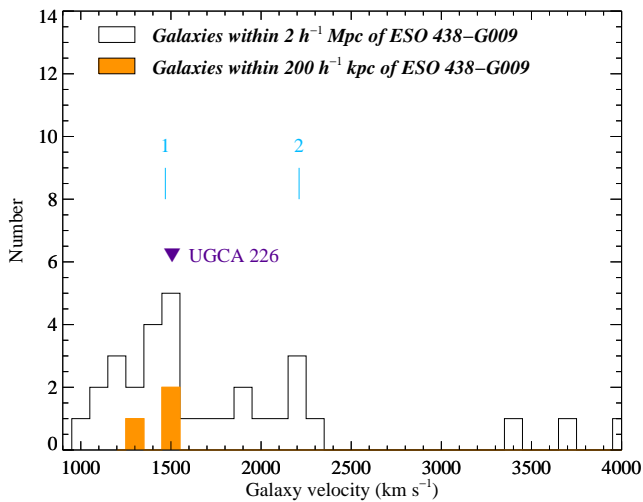}
\caption{\label{fig_es438wide} \footnotesize Distribution of all galaxies with known
redshifts within a distance of 2 and 0.2~\mpc\ of ESO 438$-$G006. 
The velocity of our selected galaxy UGCA~226 is marked with a $\blacktriangledown$.
The velocities of the \lya\
lines are coincident with the peaks in the galaxy distribution.}
\end{figure}

UGCA~226 is a highly inclined SB(s)m galaxy with a velocity of
$1507$~\kms\ lying 25.2 arcmins or 109.6~\h\ from ESO~438$-$G009.  A
DSS image of the field (Figure~\ref{fig_ES438dss}) shows the galaxy to
be relatively isolated, with no obvious companions.  ESO~438$-$G012
($1322$~\kms), 35.9 arcmins away ($\equiv 137$~\h ), is the next
nearest galaxy with a known redshift, with ESO~438$-$G010 ($1487$~\kms
) a close second, 36.3 arcmins ($\equiv 156$~\h ) away. These objects
may constitute a loose group of galaxies.  Within a 2~$h^{-1}$~Mpc
radius of the sightline there are additional galaxies at the
velocity of UGCA 226 and its neighbors (Figure~\ref{fig_es438wide}). The
nearest cataloged cluster is Abell~S0657, 29.9 arcmins away, but no
redshift information is available for that cluster. The NGC~3923 group
is 8.8 degrees away, or 3.7~\mpc\ given its velocity of 1788~\kms ---
UGCA~226 and its neighbors may lie at the outer edges of that group.

Figure~\ref{fig_spectra} shows that \lya\ is detected close to the
redshift of UGCA~226, even though the S/N of the spectrum is poor (the
data are rebinned to 0.1~\AA\ pix$^{-1}$ in the figure). As a
result of the low quality of the data, and the position of the
absorption on the damped wing from Milky Way absorption, a $\chi^2$
fit of a theoretical line profile is not well constrained,
particularly blueward of the line where the flux is approaching
zero. Hence we have primarily used the red wing of the line to derive
$b\sim 30$~\kms\ and $\log N$(H~I)$\sim 14.13$ at $v=1469$~\kms\ for a
single cloud. It is possible that additional absorption exists to the
blue of this component, but without better quality data, adding
additional components is unjustified. Also visible in
Figure~\ref{fig_spectra} is complex Si~III~$\lambda 1206$ at $z=$0.0237,
0.0229 and 0.0216, close to the emission redshift of the AGN
($z_{\rm{em}}=0.025$). Not shown in the figure is the corresponding \lya\
absorption, which is extremely strong and indicative of a Broad
Absorption Line (BAL) feature.

The only nearby galaxy with a known redshift similar to that of a strong
\lya\ line at 2211~\kms\ (component 2) is ESO 438$-$G006, an irregular
$M=-16.1$ galaxy 110 arcmins or 696~\h\ from the sightline.

This field is obviously similar to that previously described towards
PKS~1004+130 in \S\ref{sect_pks1004}, with the existence of three galaxies
all within 200~\h\ of the QSO sightline, and roughly similar luminosities
(compare entries in Table~\ref{tab_others}). In that case the absorption is
half as strong and does not show obvious complex multi-structure, although
the data is of lower S/N than that of PKS~1004+130 and the line is on the
wing of the Milky Way damped \lya\ profile. Better quality data is needed
to fully resolve whether there is a large number of components comprising
the line.

\subsection{MCG+10$-$16$-$111 \& NGC~3613/3619 \label{sect_mcg10}}

\begin{figure*}
\centerline{\psfig
{figure=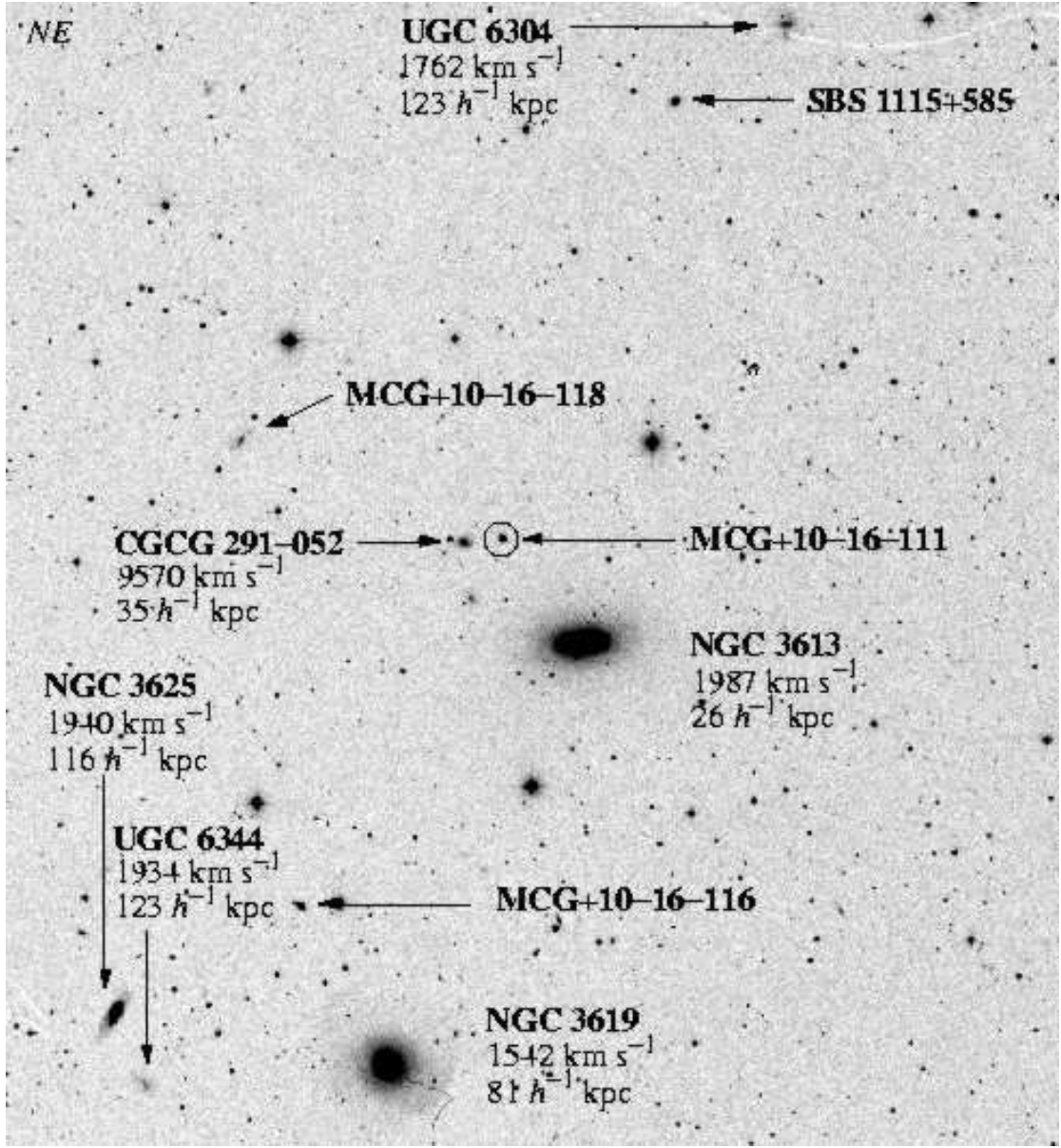,height=16cm,angle=0}}
\caption{\label{fig_MCG10dss} DSS image of the field around MCG+10$-$16$-$111
(circled). NGC~3613 ($M_B=-19.8$) lies 26~\h\ from the sightline, while
NGC~3619 ($M_B=-18.5$) lies 81~\h\ away. Other galaxies in the field are
indicated---redshifts are labelled if available.
For scale, the separation between MCG+10$-$16$-$111 and NGC~3619 is 18.2
arcmins. The image shows that the sightline to MCG+10$-$16$-$111 passes through a rich group
of galaxies. Strong \lya\ absorption is detected at the velocity of NGC~3619 and
NGC~3613, with absorption from the latter breaking into two distinct
components. There is also evidence for a broad \lya\ line at a velocity
between that of the two galaxies which may arise from hot, $10^6$~K gas. } 
\end{figure*}

\begin{figure}[th]
\hspace*{-1.0cm}
\psbox[xsize=0.6\textwidth,rotate=l]{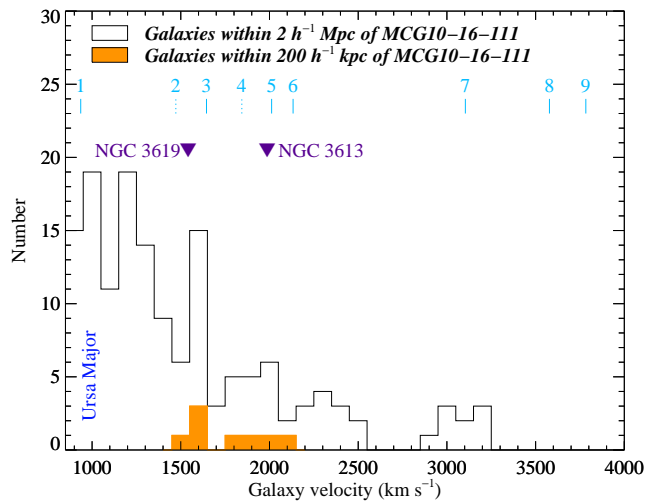}
\caption{\label{fig_MCG10dist} \footnotesize Collation of all galaxies with known
redshifts within a distance of 2 and 0.2~\mpc\ of
MCG+10$-16-111$. 
The velocities of our selected galaxies NGC~3613 and NGC~3619 are marked with a $\blacktriangledown$.
Tick marks again show the velocity of the \lya\ lines
detected. Component 3 is the strong line at the velocity of
NGC~3619, while component 2 is shown as a dotted line to emphasize that it
is the weak additional component to the blue of component 3. Components 5
and 6 are the strong lines associated with NGC~3613. Component 4 is the
weak broad feature which may have a large Doppler parameter and may be
indicative of hot intragroup gas.}
\end{figure}

\begin{figure}[th]
\psbox[xsize=0.395\textwidth]{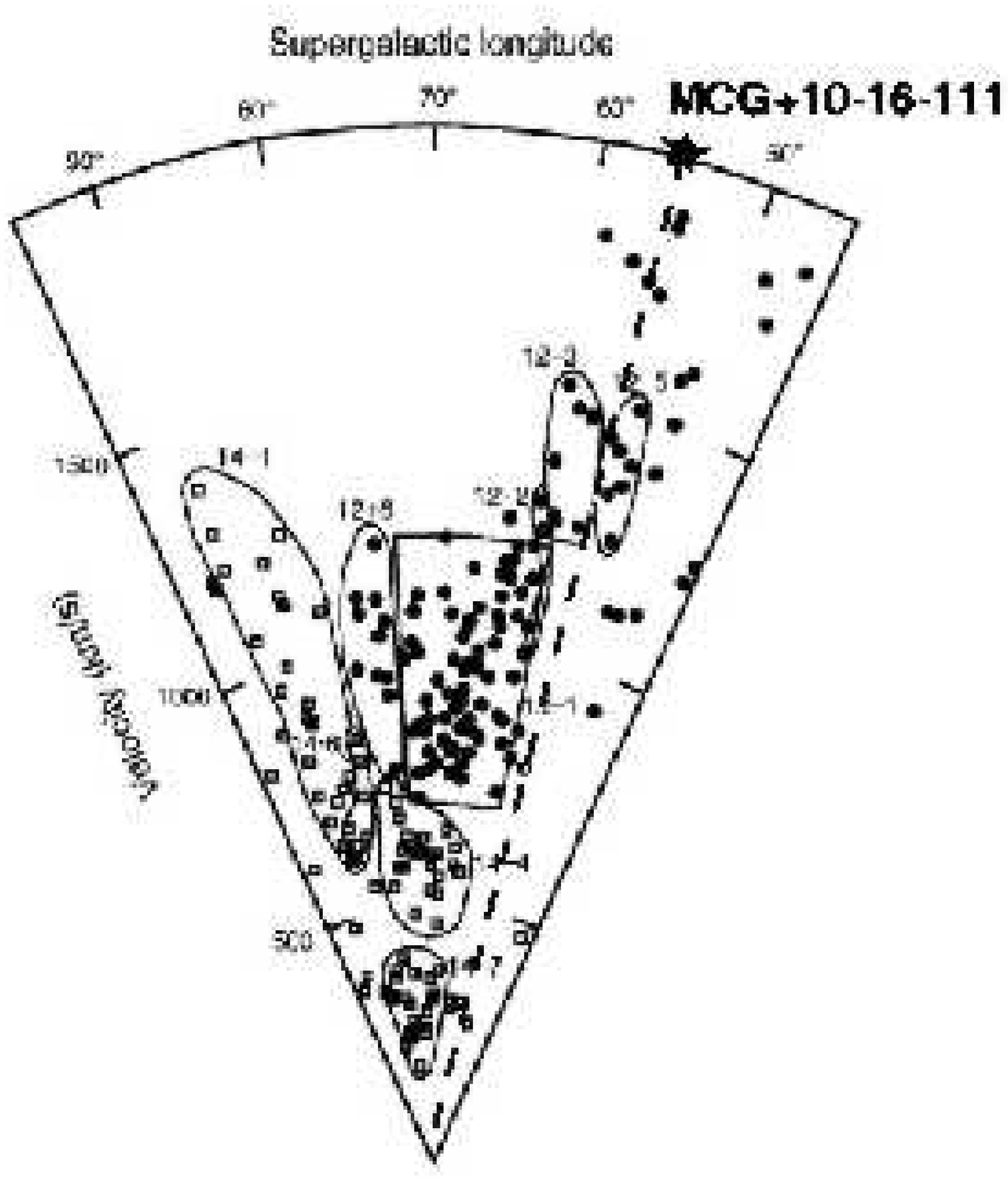}
\caption{\label{fig_ursamajor}\footnotesize Distribution of galaxies in
Ursa Major reproduced from \citet{tully96}. The figure is drawn is supergalactic
coordinates; the sightline to MCG+10$-16-111$ at SGL = $55.4^\circ$, SGB=$1.3^\circ$
is marked. The range  $-10^\circ < SGB <
15^\circ$ is included. Filled circles identify galaxies in the Ursa Major group, while
open squares are identified with galaxies in the local Coma-Sculptor Cloud.}
\end{figure}

The sightline to MCG+10$-16-$111 passes 4.5 arcmins or 25.5~\h\ from
the E6 galaxy NGC~3613, which has a velocity of $1987\pm28$~\kms .  It
also passes 18.2 arcmins or 85.3~\h\ from NGC~3619, a lenticular S0/Sa
galaxy with a velocity of $1542$~\kms . Both galaxies can be seen in a
DSS image reproduced in Figure~\ref{fig_MCG10dss}. The figure also shows
many other galaxies with velocities similar to both NGC~3619 and
NGC~3613 nearby.  Table~\ref{tab_others} lists galaxies with known
redshifts lying within 200~\h\ of the sightline. NGC~3613 and NGC~3619
obviously lie in a rich group of galaxies, and a collation of galaxies
out to 2~\mpc\ (Figure~\ref{fig_MCG10dist}) shows that NGC~3619 in
particular is part of a large over-density which arises from the outer
regions of the Ursa Major cluster. The center of Ursa Major is 10.3
degrees away, which at a velocity of $\simeq 950$~\kms\ (lower than
NGC~3613 \& NGC~3619), corresponds to 1.7~\mpc . A reproduction of the
galaxy distribution presented by \citet{tully96} in the form of a
pie-diagram is shown in Figure~\ref{fig_ursamajor}, annotated to include
the line of sight towards MCG+10$-16-$111. The velocity of NGC~3619
well matches that of the sub-groups labelled 12-3 and 12-5. Several
galaxies can also be seen at the higher velocity of NGC~3613.

At the time we selected MCG+10$-16-$111 to be observed with HST, the
redshift of the closest galaxy to the sightline on the plane of the sky,
CGCG 291$-$052, with $\rho = 1.3$ arcmins, was not known. We obtained a
spectrum of this galaxy using the ISIS spectrograph at William Herschel
Telescope on La Palma on 26-Nov-1998, and measured a velocity of
$9570$~\kms. The redshift of this galaxy has also now been listed in the
Updated Zwicky Catalog \citep{falco99} with $v=9753$~\kms, close to our
measured value. These velocities give a separation of 35~\h , but our G140M
spectrum does not cover the wavelength region where \lya\ absorption would
be expected.

Strong \lya\ lines are seen at velocities close to both NGC~3619 and
NGC~3613 (Figure~\ref{fig_spectra}). The strongest line (component 3 in
Figure~\ref{fig_spectra} and Table~\ref{tab_cols}) at 1645~\kms\ is
+103~\kms\ from NGC~3619's systemic velocity (and $-$340~\kms\ from
NGC~3613). Si~III~$\lambda 1206$ is also detected at the same velocity
as this line. Although the strong line can be well fit with a single
theoretical component, an additional weak, broad line is required at
1472~\kms\ (component 2) to fit the blue wing of component 3. 
This component is needed because there appears to be a significant
deviation in the continuum at that velocity. If this line is really composed
of only one component, then it seems to be particularly broad, with $b\geq
300$~\kms . Unfortunately, the blending of the line with the rest of the
complex makes it very hard to reliably constrain either $b$ or $N$(H~I).

Absorption is seen close to the velocity of NGC~3613 (1987~\kms ) as well,
split into two components (5 and 6) at 2012 and 2133~\kms .  There is also
need for a weak component {\it between} the strong lines (component 4)
since a clear decrement exists in the continuum between component 3 and
components 4/5.  The profile fit to the weak feature is not well
constrained, blended as it is with the strong flanking lines. Formally, we
derive $b= 101$~\kms\ if the line is a single component. If the line width
was purely a function of the gas kinetic temperature, $T$, then $T=58.8
b^2$, or $0.6\times10^{6}$~K. Such temperatures might be indicative of a hot gas cloud
cooling within the intragroup medium.

The velocities of galaxies in the Ursa Major cluster itself range from
800$-$1300~\kms, and we detect \lya\ absorption at $v=936$~\kms\
(component 1). Determining whether this arises from gas loosely bound
to the cluster, or from a cluster member much closer to the line of
sight, will require a complete redshift survey of the field.

NGC~3613 is the galaxy with the smallest separation from a QSO
sightline in our sample (26~\h ), and the field around the line of
sight has the highest density of galaxies of galaxies of all the
fields presented herein.  We can consider the absorption in two
ways. First, the total spread of absorption components $2-6$ is
660~\kms, similar to the 740~\kms\ seen towards PKS~1004+130
(\S\ref{sect_pks1004}). The fields would seem to be analogous, in that
a rich group of galaxies has produced strong, complex absorption.  The
total equivalent width of components $2-6$ is $1.1\pm0.03$~\AA , the
strongest complex in our sample.  Our original goal, however, was to
explore how gas was distributed around individual galaxies. The
detection of the strongest lines close to the velocity of the nearest,
brightest galaxies naturally leads us to view individual absorption
components as associated with individual galaxies. Indeed, the
distribution of components in velocity fits well the model of
individual galaxies surrounded by their own halos: NGC~3613 and
NGC~3619 both have systemic velocities within 150~\kms\ of the strong
absorption components (3 and 6 in Figure~\ref{fig_spectra}) while
component 4 could be ascribed to the next nearest bright ($M_B=-18.2$)
galaxy, NGC~3625, which has a velocity of 1940~\kms\ (a difference of
96~\kms\ from the \lya\ line) and is 116~\h\ from the line of sight.

Again, however, Figures~\ref{fig_MCG10dist} and \ref{fig_ursamajor} warn
that such associations may be coincidental. There is no doubt that
component 3 of the absorption complex is also `associated' with a strong
and (apparently) narrow over-density of galaxies within a 2~\mpc\ radius of
the line of sight. In this case we know the large-scale structure of the
galaxy distribution, namely the Ursa Major cluster.  There is a hint of a
similar over-density at the velocity of component 5, although here the peak
is small and poorly defined. It may be that some or all of the absorption
seen actually arises from intracluster gas.  The detection of a broad \lya\
component (number 4) may also support such an idea.  If true, the
coincidence in velocity between absorbing gas and the two individual bright
galaxies suggests that both gas and stars must closely follow the same dark
matter potentials.

\subsection{PG1149$-$110 \& NGC~3942 \label{sect_pg1149}}

\begin{figure}
\vspace*{-2cm}\centerline{\psfig
{figure=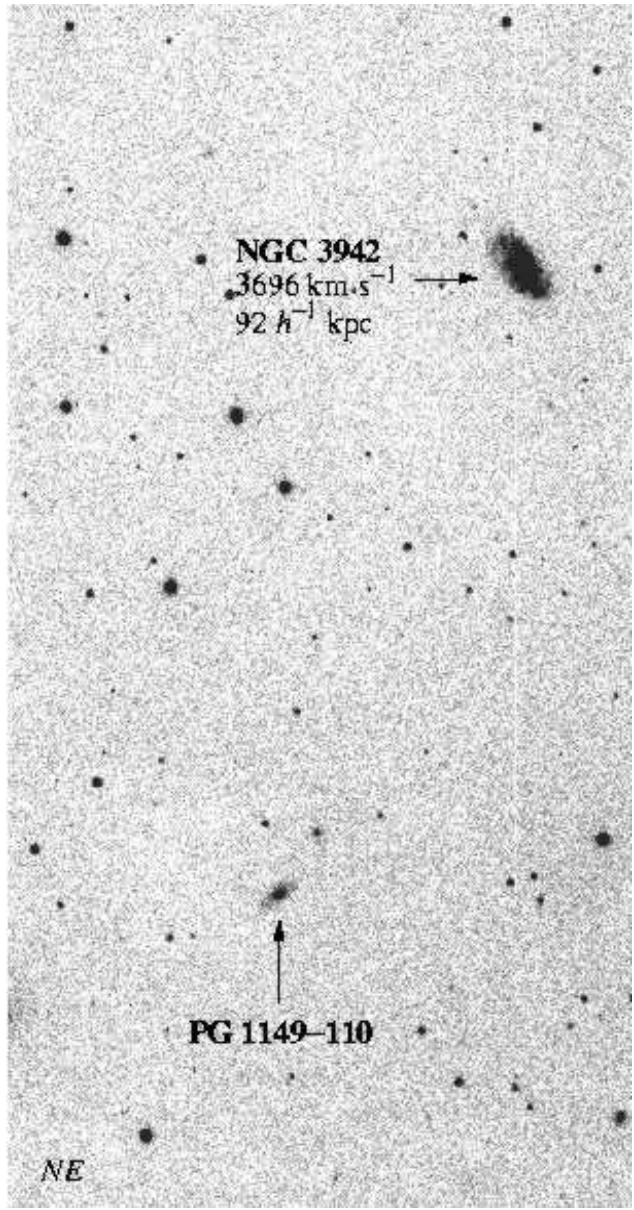,height=16cm,angle=0}}
\caption{\label{fig_PG1149wfc} \footnotesize 
INT WFC image of the field around
PG~1149$-$110. NGC~3942 ($M_B=-19.1$) is 92~\h\ away from the
sightline. For scale, the separation between QSO and galaxy is 8.7
arcmins. \lya\ absorption is detected at a velocity of 3716~\kms .}
\end{figure}

\begin{figure}[th]
\hspace*{-1.0cm}
\psbox[xsize=0.6\textwidth,rotate=l]{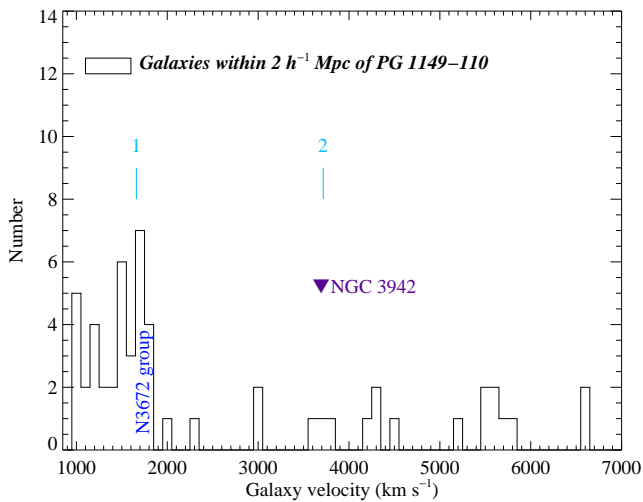}
\caption{\label{fig_pg1149wide}\footnotesize Distribution of all galaxies with known
redshifts within a distance of 2~\mpc\ of PG~1149$-$110. 
The velocity of our selected galaxy NGC~3942 is marked with a $\blacktriangledown$. }
\end{figure}

The sightline to PG1149$-$110 passes 8.7 arcmins or 92~\h\ from the
SAB(rs)c galaxy NGC~3942 which has a velocity of $3696$~\kms .
Figure~\ref{fig_PG1149wfc} shows a 200 sec $B$-band INT WFC image of the pair
taken 24-Nov-1998.  The next nearest galaxy with a known redshift is
MCG$-02-30-033$, and, with a velocity much lower than NGC~3942 (1271~\kms )
and an angular separation of 69 arcmins, is 253~\h\ from the sightline.
This galaxy appears to form part of a group of galaxies at a velocity $\sim
1500$~\kms\ (see Figure~\ref{fig_pg1149wide}). The nearest galaxy at a
velocity similar to NGC~3942 is $\simeq\:1\:h^{-1}$~Mpc away, suggesting
that NGC~3942 is relatively isolated, although this may simply be that few
redshifts have been obtained in the region beyond 3000~\kms . 

The S/N of the STIS spectrum obtained is poor, and the data are
rebinned by a factor of two to 0.1~\AA\ pix$^{-1}$ in
Figure~\ref{fig_spectra}. Nevertheless, moderately strong absorption is
detected at the redshift of NGC~3942 (component 2 in
Figure~\ref{fig_spectra} and Table~\ref{tab_cols}).

The nearest cataloged collection of galaxies is the NGC~3672 group,
which lies 7.2$^\circ$ away from the QSO sightline. At the velocity of
NGC~3672 itself (1862~\kms ) this separation corresponds to 2.3~\mpc
. It is likely that galaxies shown in Figure~\ref{fig_pg1149wide}
with velocities $\sim 1700$~\kms\ are part of the NGC~3672 group.
\lya\ absorption is also detected at 1660~\kms\ in the G140M spectrum
(component 1 in Figure~\ref{fig_spectra}), and it may be that this
line is associated with either intragroup gas or with an as yet
unidentified group member. Unfortunately, the spectrum is particularly
noisy at that point, and we cannot derive a reliable column
density. The previous sections have indicated that in several cases
absorption detected at galaxy redshifts 
splits into multicomponents spread over
several hundred ~\kms.  We are unable to tell from the spectrum of
PG~1149$-$110 how complex the \lya\ line is, but we note that there does
appear to be a depression in the continuum to the blue of the line
(between 3100 and 3500~\kms in Figure~\ref{fig_spectra}). It would be
presumptuous to claim this feature as detection of absorption, but
higher quality data would be useful in determining if the system
actually comprises additional weaker absorption components spanning
several hundred ~\kms .

\subsection{PG~1341+258 \& G1341+2555}

The sightline to PG~1341+258 passes by the North Galactic Pole, 1.9 arcmins
from G1341+2555, although it avoids the high density regions of the Virgo
and Ursa Major clusters.  Included in the QSO-galaxy compilation of
\citet{monk86}, the redshift of G1341+2555 was originally listed by
\citet{wrongz} as 5802~\kms , giving a QSO-galaxy separation of 31~\h
. However, since our HST program was completed, a new redshift measurement
has been published by \citet{grogin00} [and included in the Updated Zwicky
Catalog \citep{falco99}]. The new value is unambiguously measured to be
$v_{\rm{gal}}=14,500$~\kms. Unfortunately, although the impact parameter
between QSO and galaxy is still of interest, $\rho = 75$~\h , our STIS
observations do not cover the wavelength region where \lya\ is expected
(1275\AA ). Although the small impact parameter implied from the original
redshift made the QSO-galaxy pair well-suited for our program, the new
measurement means we must exclude the pair from further
analysis. However, we do reproduce the spectrum of the sightline in
Figure~\ref{fig_spectra}, since we will use the data in
\S\ref{sect_lss}. We see a single, weak  \lya\ line at a velocity of
1425~\kms .

\subsection{Q1831+731 \& NGC 6654/NGC~6654A}

\begin{figure}
\centerline{\psfig
{figure=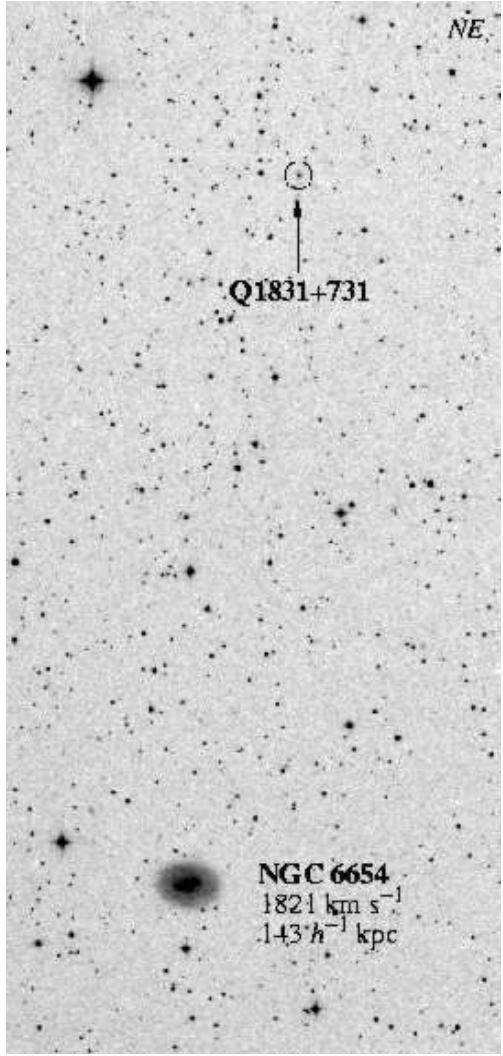,height=14cm,angle=0}}
\caption{\label{fig_Q1831dss}  DSS image of the field around
Q1831+731, with NGC~6654 ($M_B=-18.9$) lying 143~\h\ away from the QSO line of sight. For scale, the
separation between QSO and galaxy is 27.2 arcmins. Weak \lya\ is detected
at 1938~\kms, but the line is very broad, and may indicate the presence of
a hot gas with a temperature $\sim10^6$~K. }
\end{figure}

\begin{figure}
\vspace*{-0.5cm}\centerline{\psfig
{figure=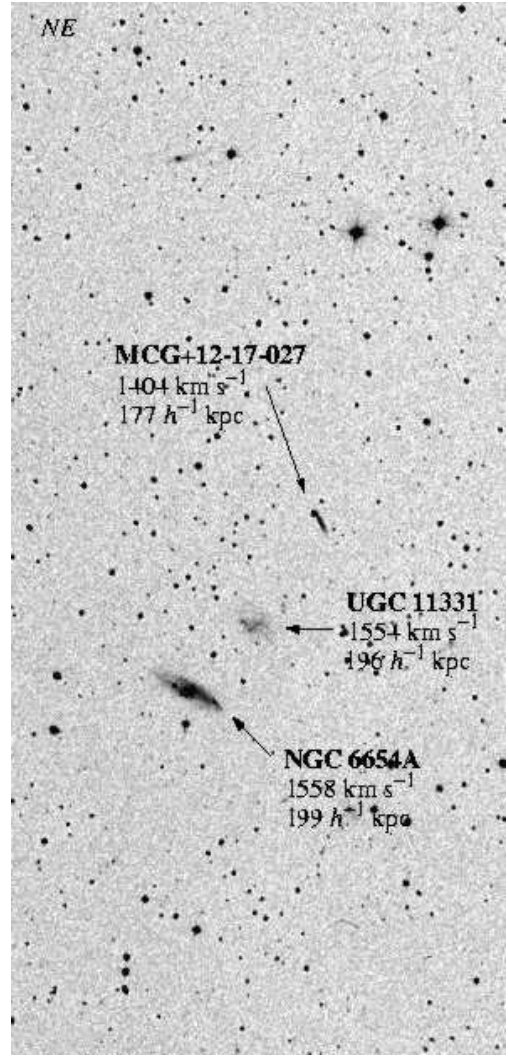,height=14cm,angle=0}}
\caption{\label{fig_NGC6654Adss} DSS image of the field around NGC~6654A
($M_B=-18.5$) which lies 199~\h\ from Q1831+731 (the QSO is not shown here ---
see Fig.~\ref{fig_Q1831dss}). NGC~6654A is the brighter member of a compact group
of at least three galaxies. Weak \lya\ absorption is detected at a velocity of
1536~\kms .}
\end{figure}

\begin{figure}[th]
\hspace*{-1.0cm}
\psbox[xsize=0.6\textwidth,rotate=l]{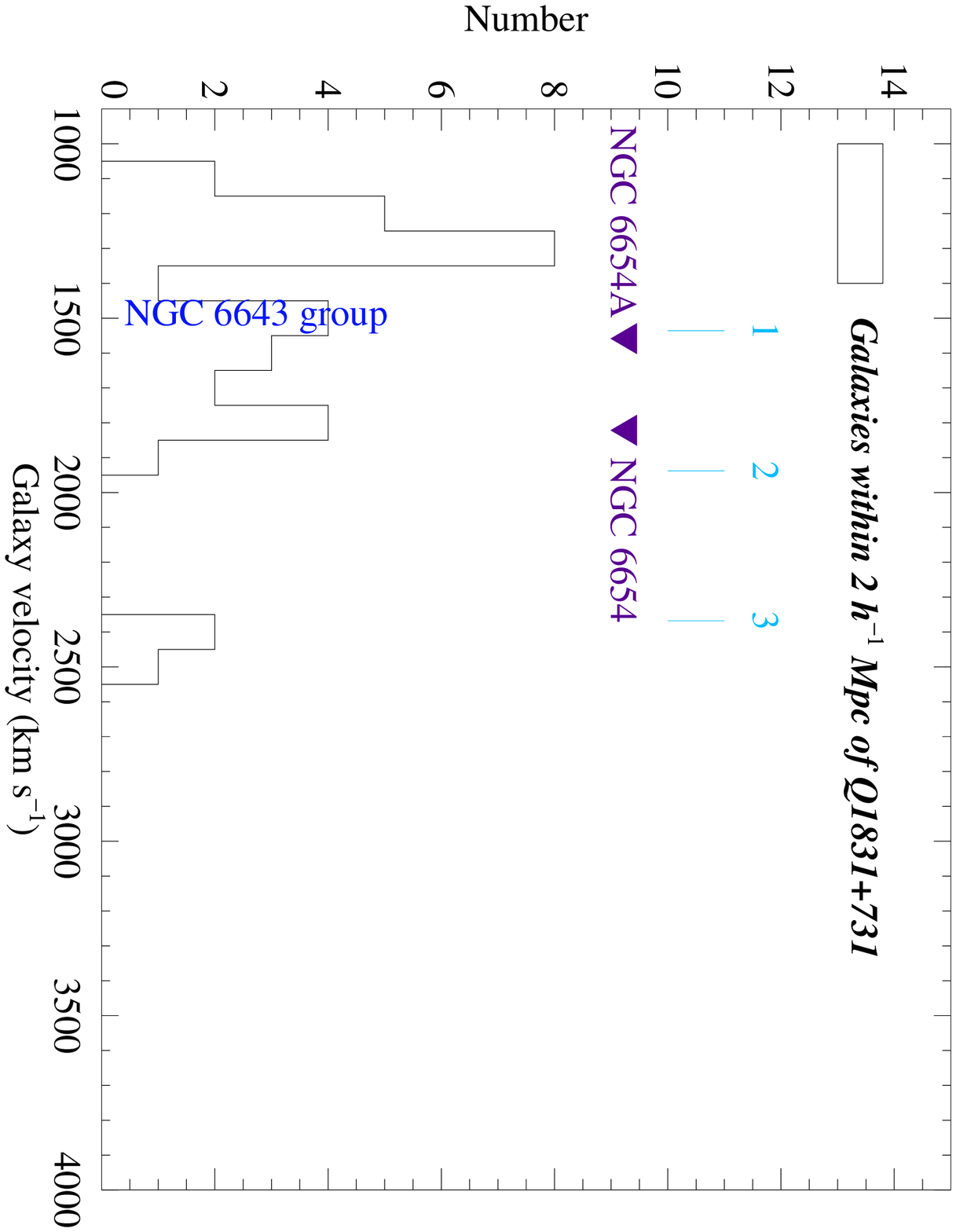}
\caption{\label{fig_q1831wide}\footnotesize Distribution of all galaxies with known
redshifts within a distance of 2~\mpc\ of Q1831+731. 
The velocities of our selected galaxies NGC~6654 and NGC~6654A are marked with a $\blacktriangledown$.
There is some indication of a
coincidence between an over-density of galaxies (including NGC~6654 and NGC~6654A) and
\lya\ absorption.}
\end{figure}

The nearest galaxy to the sightline towards Q1831+731 is NGC~6654, an
SB(s)0/a galaxy with a velocity of 1821~\kms , which, given the impact
parameter of 27.2 arcmins, corresponds to 143~\h . A reproduction of a
DSS image of the galaxy can be seen in Figure~\ref{fig_Q1831dss}.
Further away and  at a lower velocity (1558~\kms ) lies
NGC~6654A. Figure~\ref{fig_NGC6654Adss}, also taken from the DSS,
shows that NGC~6654A is actually the brighter member of a group of (at
least) three, with the dwarf UGC~11331 ($v=1554$~\kms ) and a fainter
galaxy MCG+12$-17-027$ ($v=1404$~\kms ) making up the group's
membership. The properties of all four galaxies are listed in
Table~\ref{tab_others}.  These latter galaxies are almost certainly
part of the NGC~6643 group (Figure~\ref{fig_q1831wide}). NGC~6643 has a
velocity of 1484~\kms, and is separated by 92 arcmins or 395~\h\ from
the line of sight.

The sightline also passes 251 arcmins from NGC 6503, a nearby galaxy
with a velocity of only 26~\kms . However, its distance from us is
thought to be $\sim 6$~Mpc \citep{bottema97,karshar}, which makes the
impact parameter $\sim 440$~kpc.  The low recession velocity also
ensures that any \lya\ absorption would be lost at the bottom of the
damped \lya\ profile of our own Galaxy.

Figure~\ref{fig_spectra} shows that three weak absorption lines are
seen near the velocities of both NGC~6654 and NGC~6654A. We are again
presented with the dilemma of whether to associate all components
with NGC~6654, the NGC~6654A group, and/or the NGC~6643 group, or
whether we should ascribe individual absorption components to
individual galaxies.  The velocity spanned by all three components is
830~\kms, again similar to the spread in velocity seen towards
PKS~1004+130 (\S\ref{sect_pks1004}) and MCG+10$-$16$-$111
(\S\ref{sect_mcg10}), although the lines are considerably weaker.  The
coincidence in velocity between component 1 and the NGC~6643 group, as
shown in Figure~\ref{fig_q1831wide}, also suggests that absorption
could be from intragroup gas. The Doppler width of component 2 is 
unusually wide, with $b=148\pm20$~\kms, 3$-$5 times larger than the usual
values seen in higher resolution GHRS data
\citep[e.g.][]{penton00}. Although the line could be composed of more
than a single component, if absorption arises from a single cloud,
the inferred kinetic temperature is $\simeq 1.3\pm0.3\times
10^6$~K, a value which might reflect gas cooling in a cluster.

\subsection{Mrk~1048 \& NGC~988}

\begin{figure}
\vspace*{-2cm}\centerline{\psfig
{figure=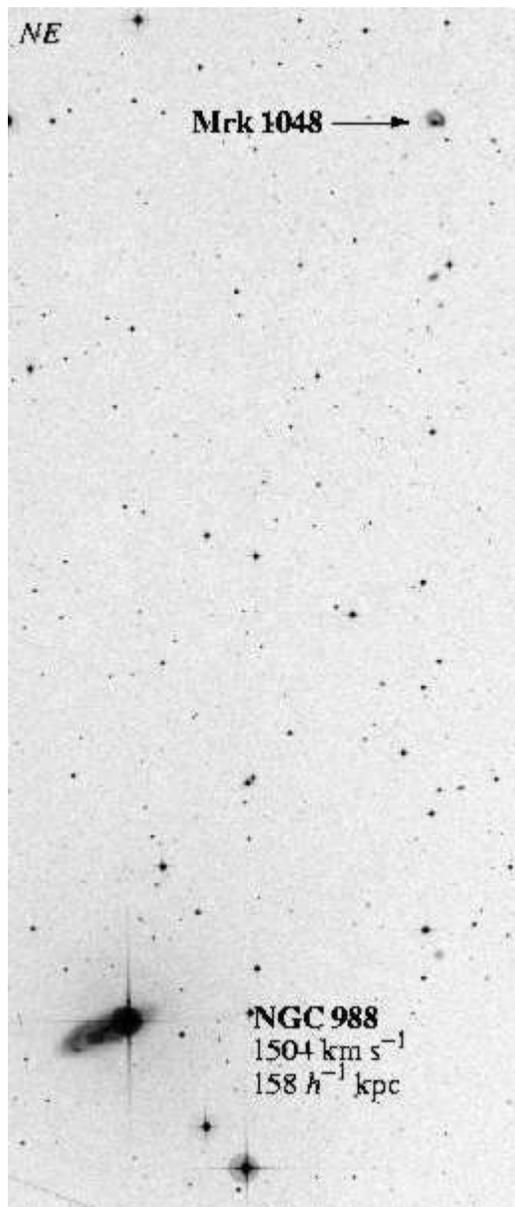,height=16cm,angle=0}}
\caption{\label{fig_NGC988dss}\footnotesize DSS image of the
field around Mrk~1048, which lies 158~\h\  from NGC~988 ($M_B =
-20.0$). For scale, the separation between AGN and NGC~988 is 36.3
arcmins. Weak \lya\ absorption is found towards Mrk~1048 at $v=1913$ and
2183~\kms . The first of these lines appears quite broad, with a doppler
parameter of 162~\kms , and may indicate
absorption by gas with a temperature of $\sim 2\times10^6$~K.}
\end{figure}

\begin{figure}[th]
\hspace*{-1.0cm}
\psbox[xsize=0.6\textwidth,rotate=l]{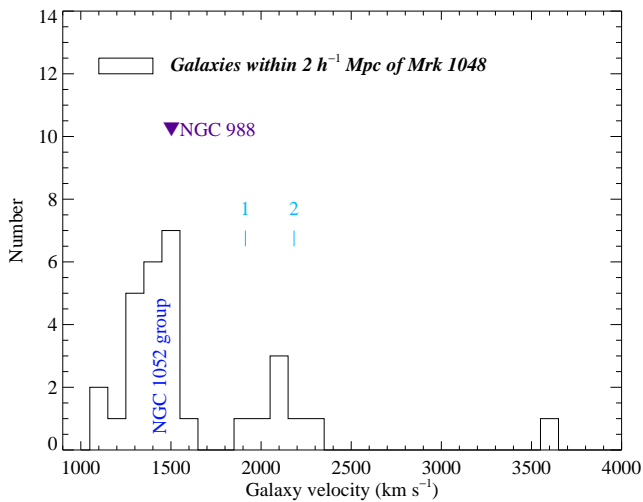}
\caption{\label{fig_mrk1048wide}\footnotesize Collation of all galaxies with known
redshifts within a distance of 2~\mpc\ of the sightline towards
Mrk~1048. Most of the galaxies are associated with the group centered on
NGC~1052, which itself has a velocity of 1473~\kms\ and is 429~\h\
from the sightline. The velocity of our selected galaxy NGC~988 is marked with a $\blacktriangledown$. }
\end{figure}

The sightline to Mrk~1048 passes 36.3 arcmins or 158~\h\ from NGC~988, an
SB(s)cd with a velocity of 1504~\kms .  Inspection of a DSS image around
Mrk~1048 (Figure~\ref{fig_NGC988dss}) shows no obvious bright galaxies closer
to the sightline than NGC~988, suggesting that the galaxy is relatively
isolated. However, \citet{vanG86} suggest that NGC~988 may be part of the
NGC~1052 group, which, combined with the NGC~1069 group, makes up the
Cetus~I cloud. NGC~1052 is 100.9 arcmins from the
sightline, or 429~\h\ at its velocity of 1473~\kms . Hence the
sightline to Mrk~1048 intercepts the group (whether or not NGC~988 is a
member) at this impact parameter assuming NGC~1052 marks the group's
center.  Within 1~\mpc\ of the sightline there are 21 galaxies with known
redshifts, of which 14 lie within $^{+59}_{-232}$~\kms\ of NGC~1052
(Figure~\ref{fig_mrk1048wide}).

Weak \lya\ absorption is detected at velocities of 1913 and 2182~\kms,
+409 and +678~\kms\ redward of NGC~988's systemic velocity. Profile
fitting of the first of these components is difficult since the line
is weak and broad, and lies on a region of the spectrum where the flux
is declining rapidly due to Milky Way \lya\ absorption.  Nevertheless,
we derive a value for the Doppler parameter of $b=162\pm33$~\kms\ for
 component 1 (Table~\ref{tab_cols})
and if the line is composed of only one component, it
may represent absorption from hot gas with a kinetic temperature of
$\simeq 1.5\pm0.6\times10^6$~K.

The \lya\ lines towards this target are the most poorly matched
absorption features in our sample to either individual galaxies or to
the groups of galaxies.  Compared to the overall distribution of
galaxies with known redshifts within 2~\mpc\ of the sightline
(Figure~\ref{fig_mrk1048wide}), the first absorption line at
1913~\kms\ is close to, but certainly not coincident with, the
velocity of the NGC~1052 group. In fact, both absorption lines appear
to lie in velocity between the NGC~1052 group and a loose collection
of galaxies comprising of Mrk~1042, DDO023, Mrk~1039 and VV~525, some
800~\h\ from the sightline. The only evidence that suggests any of the
gas might be intragroup gas is the large Doppler parameter of
component 1.  The detection of absorption at a velocity close to
NGC~988 makes it possible that the absorber is related to the galaxy,
but the 400~\kms\ velocity difference is quite large, and would
probably have to be explained by material infalling or outflowing from
the galaxy.

    \section{Correlations and Interpretations \label{sect_corrs}}

The previous section indicates that we find absorption within
500~\kms\ of all eight galaxies studied, with individual \lya\ lines appearing
to cluster at the same  velocities as the galaxies.  The original
intention of this program was to establish how the physical properties
of a \lya\ absorbing cloud correlate with those of a particular
galaxy.  However, to make such a connection, we immediately encounter
the problem of how to associate an absorbing cloud with an individual
galaxy. In our survey, there are several cases where absorption arises
in several components close to (but not necessarily coincident with)
the systemic velocity of a galaxy. The first problem is deciding which
components to associate with the galaxy. The situation is further
complicated if there is more than one galaxy close to the line of
sight (e.g., NGC~3616/3619 and NGC~6654/6654A) and at similar
velocities, since absorption assigned to one galaxy might be more
properly associated with another.

For example, we could simply take as `associated' the absorption
component nearest in velocity with that of a probed galaxy. The
consequences of this would be that all the substructure seen in the
lines towards PKS~1004+130 (components 3,4,6 and 7 in
Figure~\ref{fig_spectra}) would not be counted, which would likely be
an inadequate description of the absorbing gas along the
sightline. Also, towards MCG+10$-16-111$, component 5 would be
associated with NGC~3613, although it is actually the weaker of the
two lines (components 5 and 6) forming the absorption structure at
$\sim 2100$~\kms . Hence this method seems
unsatisfactory. Alternatively, we could count all absorption
components within a velocity interval $\pm \Delta v$ of the systemic
velocity of the galaxy, taking $\Delta v$ to be, say, some value
between that expected for gas co-rotating in a disk ($\sim
100-200$~\kms , depending on galaxy inclination) and the dispersion of
a fairly rich galaxy group ($\sim 500$~\kms ). Such a scheme would certainly
include all five components clustered around UGC~5454's velocity
towards PKS~1004+130 for example. However, rigorous application of the
counting scheme in cases where more than one galaxy is present would
lead to a situation of `double counting' of absorption
lines. Components 1 \& 2 in Figure~\ref{fig_spectra} towards Q1831+731
would have to each be counted for both NGC~6654 and NGC~6654A, since
both lines fall within $\pm 400$~\kms of each galaxy, and the value of
$\Delta v$ would need to be at least this high because of the
difference in velocity between absorber and galaxy.  Similarly, the
strong absorption component 3 towards MCG+10$-16-111$ would have to be
counted once for NGC~3619 and again for NGC~3613, since the absorption
lies $+103$~\kms\ from NGC~3619 and $-342$~\kms\ from NGC~3613.

Of course, such double counting could be overcome by making the chosen
$\Delta v$ sufficiently small. However, at some point, $\Delta v$ is
so small that only one line is counted, and we have returned to the
first suggested method of counting absorption lines by selecting
only those closest in velocity to a galaxy.

In reality, deciding which method to use (or, equivalently, deciding
how small $\Delta v$ should be) almost {\it presupposes} some model of
how \lya\ clouds are associated with galaxies. If the underlying
assumption is that a galaxy and absorber are associated on a
one-to-one basis, e.g. as a result of absorption from a co-rotating
disk or halo, then it would make more sense to count only the
component nearest in velocity.  Conversely, choosing absorption lines
within a certain velocity range --- at least when more than one galaxy
is present --- would be a method chosen if the gas is assumed to be
part of the gravitational potential that a galaxy group shares. 

It is actually easier to consider whether our results are consistent
with those seen at higher redshift and to examine how our results
would be interpreted in higher-$z$ surveys by degrading our STIS data to
the lower resolution of FOS data. We perform such an experiment in the
following section.

\subsection{Degrading STIS data to FOS resolutions \label{sect_fos}}

\begin{figure}
\vspace*{-1cm}\centerline{\psfig
{figure=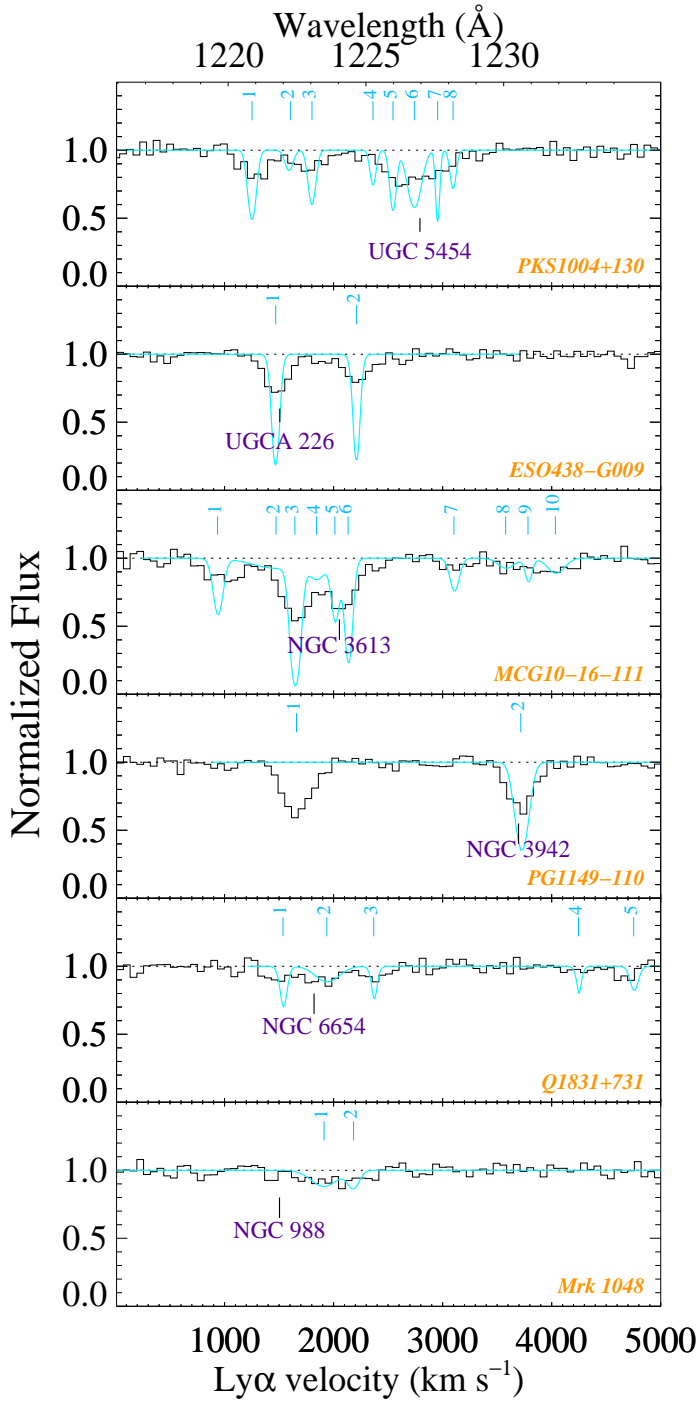,height=19cm,angle=0}}
\caption{\label{fig_simplot}\footnotesize A simulation of how the  data presented in
Figure~\ref{fig_spectra} would appear 
if degraded to the resolution of the FOS. Only the galaxies nearest to the probe line of sight 
are labelled. Grey lines represent the theoretical profiles derived
from the STIS data. The S/N of the data has been set to be 30 per pixel.}
\end{figure}

To synthesize how our QSO/AGN spectra would appear if obtained with the
FOS, we have taken the profile fits to the normalized data and convolved
the spectra with a Gaussian of FHWM of 1.0~\AA , which approximates the LSF
of the G130H FOS grating. \lya\ lines observed at higher redshift would
have been observed with all three available UV medium-resolution gratings,
but the resolutions of all the gratings were similar, $R\simeq 1300$ or
230~\kms . Post-COSTAR, the LSFs of FOS gratings show some departures
from Voigt profiles at high S/N, but no theoretical LSFs are available
\citep{Keyes97}, and the use of a Gaussian is likely sufficient for this
exercise. The convolved spectra have then been rebinned to the FOS G130H
dispersion of 0.25~\AA~pix$^{-1}$. The resulting spectra are shown in
Figure~\ref{fig_simplot}. This figure differs from the original data
presented in Figure~\ref{fig_spectra} in that we have removed the Galactic
damped \lya\ profile in order to better show the lower resolution
features. The original (STIS-resolution) theoretical line profiles are
superimposed on the plot to show how the data originally appeared. The
velocity of the nearest galaxies to the line of sight is also marked.

In these lower resolution data, how would the equivalent widths of \lya\
lines be measured and which galaxies would the lines be associated with?
LBTW and CLWB take the galaxy nearest to the sightline as being
responsible for \lya\ absorption when more than one is detected at similar
redshifts.  UGC~5454 would be taken to be the nearest galaxy to the
sightline of PKS~1004+130; the individual components within the complex
profile discussed in \S\ref{sect_pks1004} would be unresolved. We assume
that only this broad line would be identified as the absorption system, and
that weaker lines (components 1 and 2) would not be included in a
measure of the equivalent width.  The values of the total equivalent
width, $W_T$ and $\rho$ for UGCA~226
towards ESO~438$-$G009, and NGC~3942 towards PG~1149-110, would be
unaffected.  The absorption system towards MCG+10$-$16$-111$ would appear
as two lines, but most likely counted as one system, with NGC~3613
identified as the absorbing galaxy. Again, we adopt $W_T$ for only the two
lines, and do not include a contribution from component 1.  It is somewhat
harder to decide which, if any, of the components towards Q1831+731 and
Mrk~1048 would be detected in FOS data, since the lines are weak, and their
detection would be much more sensitive to the exact S/N of the data. From
our simulation, although the individual components towards Q1831+731 are,
in principle, resolved from each other, at their respective velocities, (and
given the S/N adopted here) they blend to form a single broad line.
Similar, for Mrk~1048, the two components would be blended together to form
a shallow, single broad line.  For both cases, therefore, we take the
value of $W_T$ that would be measured to be sum of the equivalent widths of
all the components seen at higher resolution. As we shall see, taking these
sums as upper limits (if only one of two components were detected) does not
affect our conclusions.

Looking at Figure~\ref{fig_simplot}, it seems clear that with these
assignments, we are effectively summing equivalent widths of
components over a range of $\sim 1000$~\kms . The only exception is
towards ESO~438$-$G009, where components 1 and 2 are separated by
740~\kms, but we have `associated' only component 1 with UGCA~226. Let
us formalize the associations and take 1000~\kms\ as the range over
which we count `one' line, but take only the nearest galaxy to the
sightline as being associated.  We summarize which components have
been combined in Table~\ref{tab_full}.  How then do the total
equivalent widths, $W_T$, correlate with $\rho$?

\begin{figure}
\vspace*{-.5cm}\centerline{\psfig
{figure=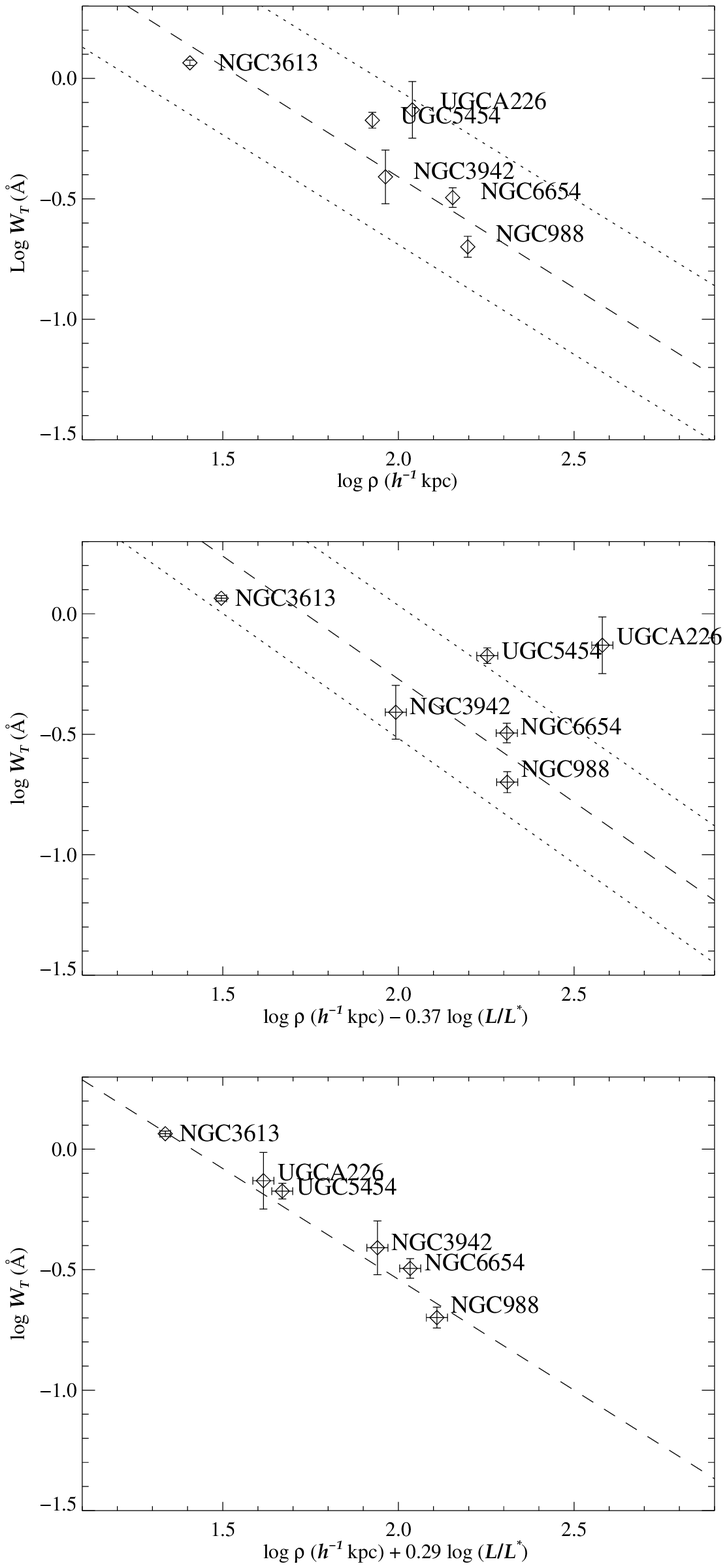,height=18cm,angle=0}}
\caption{\label{fig_corrs1} \footnotesize
Top: Plot of total equivalent width, $W_T$, against QSO-galaxy impact
parameter, $\rho$, for galaxies in our sample which are closest to the
probe line of sight. Here, $W_T$ represents the sum of equivalent
widths from all components within $\pm 500$~\kms\ of the center of the
line complex seen in Figure~\ref{fig_simplot}.  Middle: Plot of $W_T$
vs.~$\rho$ corrected for galaxy luminosity as prescribed by CLWB. In
both panels, the relationships found by these authors for
higher-redshift absorbers are shown as dashed lines, with the
approximate 1$\sigma$ scatter shown by dotted lines.  Bottom: Plot of
$W_T$ vs.~$\rho$ corrected for galaxy luminosity in a way which
minimizes $\chi^2$ for our data points and equation 1.
}
\end{figure}

The results are shown in Figure~\ref{fig_corrs1}. The plot of \lya\
line equivalent width, $W_T$, against line-of-sight impact parameter,
$\log \rho$, is shown top of the Figure, along with the relationship
found by CLWB for higher redshift \lya\ absorbers, $\log W =
-0.93\log\rho + $const.  Also plotted is the typical dispersion in
their measurements with dashed lines (these represent $\pm 68$\% of
their data points either side of their best fit).  We find that the
points lie well within the envelope of points found for high-redshift
systems, although statistically, Spearman and Kendall rank-order
correlation tests show an anti-correlation at a significance level of
only 2.0$\sigma$ each.

CLWB also found a relationship between \lya\ equivalent width, impact
parameter, and galaxy luminosity of the form

\begin{equation}
\log W = -\alpha \log \rho + \beta \log L + {\rm{const}}
\end{equation}

\noindent
with

\[
\alpha = 1.02,\:\:\beta = 0.37\hspace*{1cm} {\rm{(CLWB)}}
\]

\noindent that is, they found that brighter galaxies have larger
halos, or put another way, that brighter galaxies show stronger \lya\
equivalent widths at a given radius.  In the middle panel of
Figure~\ref{fig_corrs1}, we plot $W_T$ against $\log\rho$ for the
galaxies in our sample, corrected by a factor of $-0.37\log (L/L^*)$,
[assuming the absolute magnitude of an $L^*$ galaxy to be
$M_B^*=-19.7$ \citep{norberg01}], with CLWB's anti-correlation plotted
as a dashed line. We use the apparent magnitudes---uncorrected for
inclination---given in Table~2, since these are more likely to
represent the types of magnitudes measured by CLWB for the high
redshift galaxies.  The figure demonstrates that the dispersion in our
data points actually increases with this correction.  Formally, both a
Spearman and Kendall test find an anti-correlation at only the
0.9$\sigma$ level, lower than the significance for a correlation
without a luminosity correction.

Although the galaxies in our sample do not follow the correlation
found at higher redshift, we can still perform a simple $\chi^2$ fit of
equation 1 to our data to see if there is any dependence of $W_T$ with
galaxy luminosity. Performing such a test, we find

\[
\alpha = 0.9,\:\:\beta = -0.29
\]

\noindent 
that is, the correlation is the {\it opposite} sense to that found for
the higher-redshift sample, implying that stronger \lya\ lines are
associated with {\it fainter} galaxies. We show how the correlation is
afffected by correcting by a factor $+0.29\log (L/L^*)$ in the bottom
panel of Figure~\ref{fig_corrs1}. Although the points appear to be
better correlated, statistically, a Spearman test finds
the correlation to be significant at the 2.2$\sigma$ level, little
different from the correlation without a luminosity correction
(2.0$\sigma$). A Kendall test finds
a correlation signficant at the 2.8$\sigma$ level, a   somewhat
better improvement over the 2.0$\sigma$ significance derived without a
luminosity correction. 

In summary: by summing individual \lya\ components over a width of
$\pm 500$~\kms\ of the velocity of a galaxy next to the line of sight,
and considering only the impact parameter of the nearest galaxy, we
are able to reproduce the correlation of $W$ with $\rho$ seen at
higher redshift. The correlation is weakened if we
correct for galaxy luminosity in the way found for galaxies at higher
redshift (\lya\ equivalent width being stronger around brighter
galaxies at a given radius), but we are able to improve our
correlation by correcting in the opposite manner (so that \lya\
equivalent width  is weaker around brighter galaxies at a given
radius). We discuss this apparent correlation in
\S\ref{sect_conclusions}. 

\begin{deluxetable}{lcclrcccccc}
\rotate
\tablecolumns{11}
\tablewidth{0pc} 
\tablecaption{Integrated quantities and galaxy volume densities \label{tab_full}}
\tablehead{
\colhead{ }  & \colhead{ } & \colhead{$W_T$} &
\colhead{$\sigma(W_T)$} & \colhead{ } & \colhead{ } &
\colhead{ } & \colhead{ $v_{\rm{range}}$}  & \colhead{} & \colhead{No.~of} &
\colhead{$n$}\\
\colhead{Probe}  & \colhead{$\Sigma$} & \colhead{(\AA )} &
\colhead{(\AA )} & \colhead{$\log N_T$} & \colhead{$\sigma(\log N_T)$} &
\colhead{$B_{\rm{lim}}$} & \colhead{(\kms )} & \colhead{$M_{\rm{lim}} - 5\log h$} & \colhead{gals} &
\colhead{($h^3$ Mpc$^{-3}$)}\\
\colhead{(1) }  & \colhead{(2) } & \colhead{(3)} &
\colhead{(4)} & \colhead{(5) } & \colhead{(6) } &
\colhead{(7) } & \colhead{ (8)}  & \colhead{(9)} &
\colhead{(10)} & \colhead{(11)}
}
\startdata
\multicolumn{11}{c}{\lya\ complexes near pre-selected galaxies }\\
\hline
PKS 1004+130    & 4-8 & 0.67 & 0.05 & 14.15       & 0.22  & 15.0 & 2200 $-$ 3200       & $-17.5$ & 9  & $-$1.15 \\
ESO 438$-$G009  & 1-2 & 0.74 & 0.2  & 14.41       & 0.30  & 14.0 & 1340 $-$ 2340       & $-17.8$ & 7  & $-$1.26 \\
MCG+10$-16-$111 & 2-6 & 1.16 & 0.03 & 14.69       & 0.28 & 14.5 & 1400 $-$ 2400       & $-17.4$ & 12\phn & $-$1.02 \\
PG 1149$-$110   & 2   & 0.39 & 0.1  & 14.04       & 0.14 & 15.5 & 3200 $-$ 4200       & $-17.6$ & 3  & $-$1.63 \\
Q1831+731       & 1-3 & 0.32 & 0.03 & 13.77       & 0.26 & 15.0 & 1500 $-$ 2500       & $-17.0$ & 2  & $-$1.80 \\
Mrk 1048        & 1-2 & 0.20 & 0.02 & 13.60       & 0.33 & 14.5 & 1550 $-$ 2550       & $-17.5$ & 1  & $-$2.10  \\
\hline
\multicolumn{11}{c}{Other \lya\ complexes in STIS spectra}\\
\hline
PG 1341+258        & 1   & 0.12 & 0.03 & 13.40 & 0.08  & 15.0  & \phn925 $-$ 1925  & $-$16.4 & 1 & $-$2.10 \\
PKS 1004+130     & 1-3 & 0.43 & 0.02 & 13.94 & 0.28  & 15.0  & 1100 $-$ 2100 & $-$16.6 & 9 & $-$1.15 \\
MCG +10$-16-$111 & 7-10& 0.30 & 0.03 & 13.76 & 0.23  & 14.5  & 3050 $-$ 4050 & $-$18.6 & 3 & $-$1.63 \\
PG 1149$-$110    & 1   & 1.10 & 0.30 & ...   & ...   & 15.5  & 1160 $-$ 2160 & $-$16.2 & 10\phn & $-$1.10 
\tablecomments{Explanation of table entries: (1) QSO/AGN probe; 
(2)  component numbers in Table~\ref{tab_cols} which are covered in the integrated 
equivalent width; 
(3) the integrated equivalent width, equivalent to the sum of equivalent widths from the individual 
components covered, which would be measured over $v_{\rm{range}}$ 
; (4) error in the integrated equivalent width; (5) integrated column density, derived from the sum of
individual column densities from components listed in (2); (6) error in the integrated column
density; (7) adopted magnitude limit for the collated galaxies from the RC3; (8) velocity range 
over which galaxies were included when calculating $n$; 
(9) absolute magnitude of $B_{\rm{lim}}$ at the \lya\ complex redshift + 500 \kms ; 
(10) Actual number of galaxies brighter 
than $M_B=-17.5$  within $v_{\rm{range}}$; (11) resulting volume density of galaxies.  }
\enddata 
\end{deluxetable}

\subsection{Comparison with Large Scale Structure \label{sect_lss}}

We saw in \S\ref{sect_results} that by collating together all the
available information on galaxy redshifts in the fields of our
selected QSOs, we found evidence that \lya\ absorption is found
preferentially at peaks in the galaxy redshift distribution, even
though it is also possible to associate each absorption line with a
given galaxy.

We described at the beginning of \S\ref{sect_results} how we selected
galaxies within 2~\mpc\ of a QSO/AGN sightline to search for any peaks in
the redshift distribution at the same redshift as the selected
galaxies. We took all galaxies with known redshifts in the RC3 within 691
arcmins of the sightline (corresponding to 2~\mpc\ at a redshift of
1000~\kms ). This list was supplemented with any other galaxies from the
NED. Since the RC3 is relatively complete in magnitude however, it should
be possible to be more quantitative about the density of galaxies along the
QSO line of sight. In principle, if we know the redshifts of all
galaxies within a specified radius, measured to some limiting galaxy
magnitude, we should be able to calculate directly the volume density of
galaxies at the same redshift as the probed galaxy.

We first consider the possible errors in the RC3
magnitudes. \citet{RC3} discuss in some detail the likely errors in
the observed magnitudes, considering that the measurements are made
from many different sources. In the end, they conclude that the mean
error is in the range $0.2-0.3$ mags.  Obviously, we would like to use
magnitudes drawn from a heterogeneous data set, but until all-sky
surveys such as the Sloan Digital Sky Survey (SDSS) are complete, we
do not have such data. If \citet{RC3}'s estimate of the magnitudes are
correct, however, then their values are probably sufficiently accurate
for the purposes of this analysis.

Our initial objective is to determine how complete we might be to a given
magnitude limit if we are to use  the RC3 as a magnitude limited sample.
We construct the observed $B$-band
distribution of galaxies for all galaxies within 691 arcmin ($\equiv
2$~\mpc\
for $cz=1000$~\kms ) of each QSO
line of sight. For each field, less than 2\% of galaxies in the RC3 do not
have a measured magnitude. We have used total $B$-band magnitudes ($B_T$)
when available, otherwise we have used photographic $B$-band
magnitudes ($B_{\rm{mag}}$). We then produce a histogram of the number of
galaxies in a 0.5 magnitude interval per unit area. These are shown in the
top panels of Figure~\ref{fig_mags}.  Also plotted are the galaxy number
counts taken from the SDSS Commissioning data
\citep{yasuda01}. As we found previously, the field around Q1831+731 does not
appear to have a high density of galaxies with known redshifts. Examination
of the $B$-band distribution 
suggests that this is because there is a genuine under-density of galaxies
in the field. Conversely, we know that the sightline around
MCG+10$-$16$-$111 shows the highest density of galaxies of all the
fields presented here,
and that too is reflected in the magnitude distribution,
with an excess of bright galaxies over the average found in the SDSS
data. The magnitude distribution towards all the other fields agree with
the result from the SDSS quite well.

\begin{figure}
\hspace*{-1.0cm}\psfig{figure=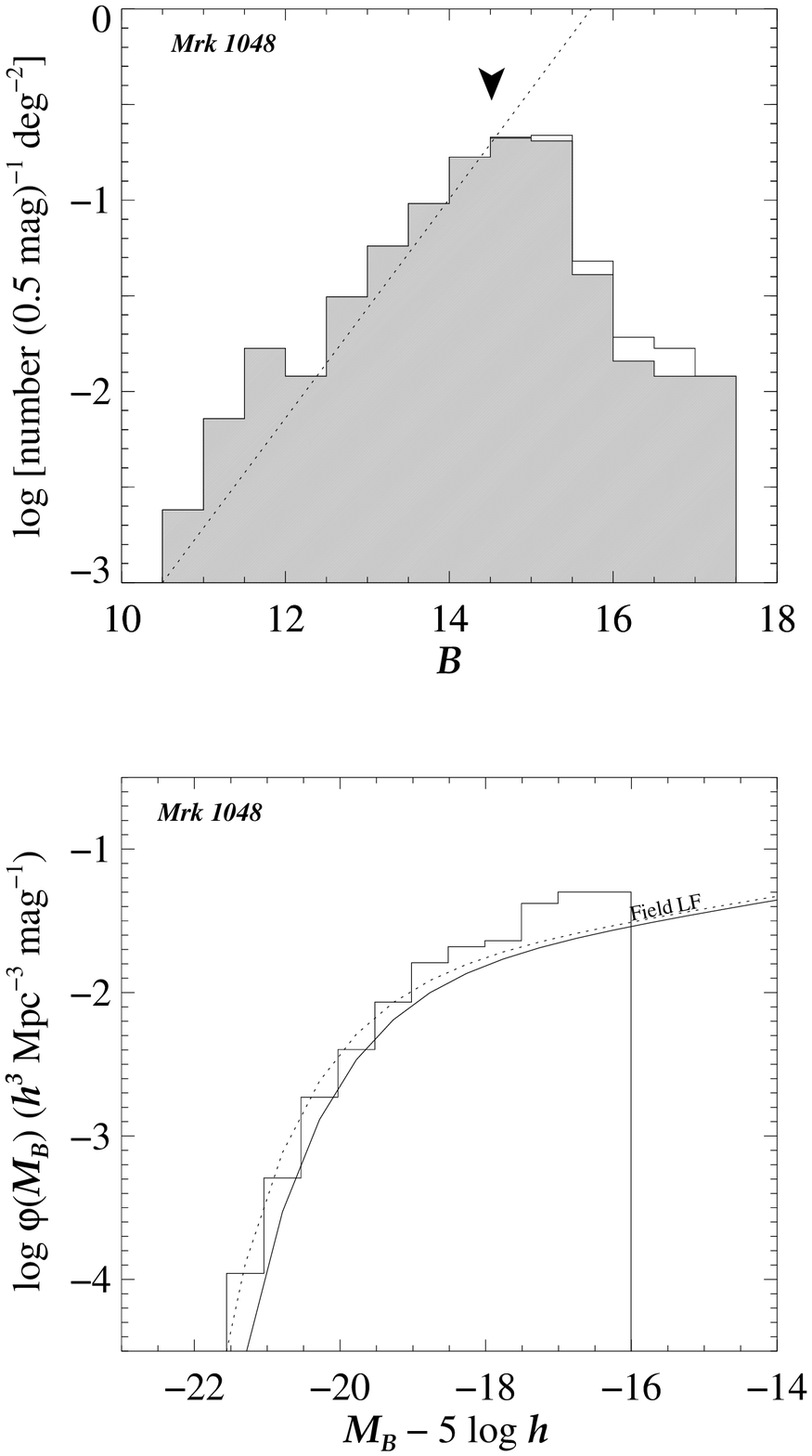,height=9cm,angle=0}\vspace*{-9cm}
\hspace*{4cm}{\psfig{figure=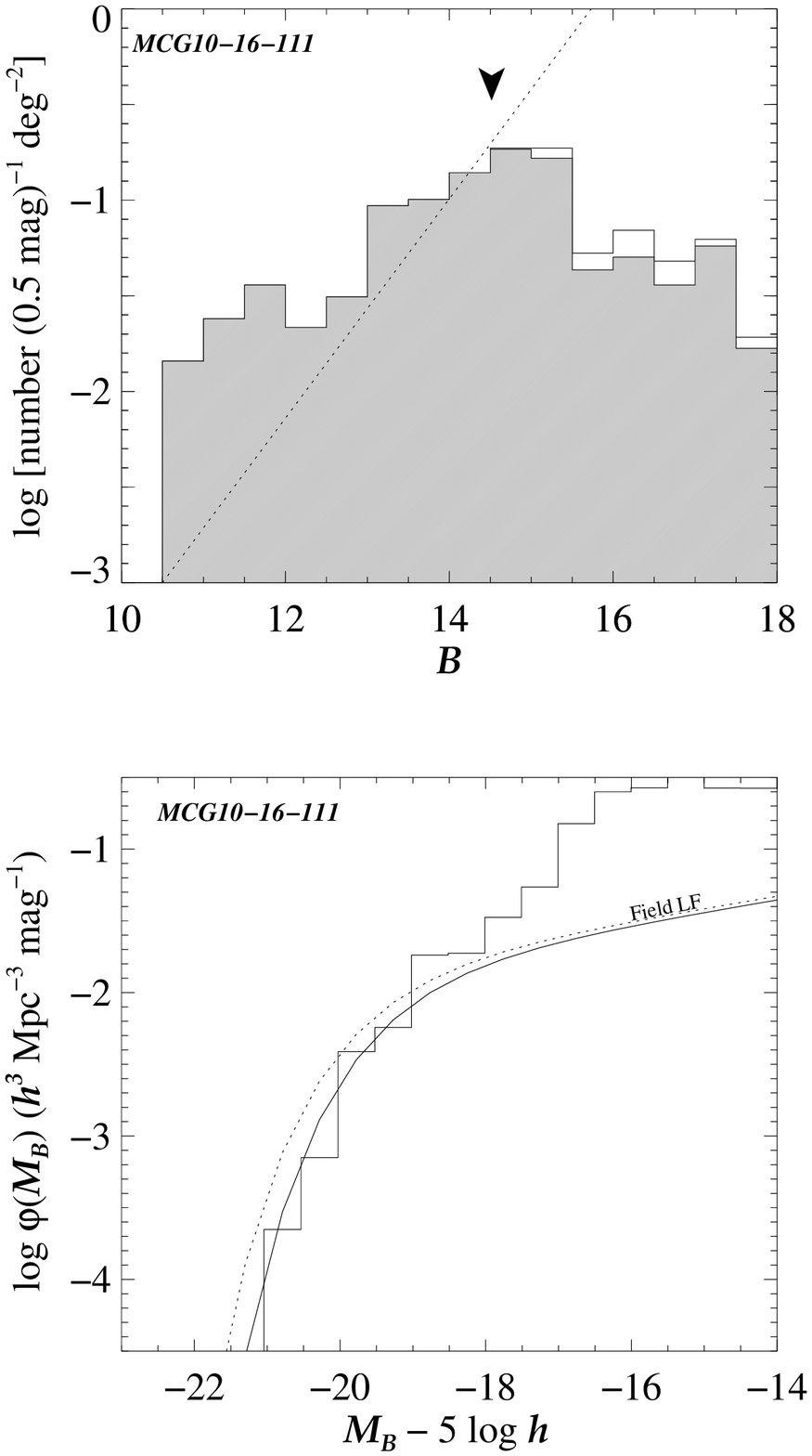,height=9cm,angle=0}}

\hspace*{-1.0cm}\psfig{figure=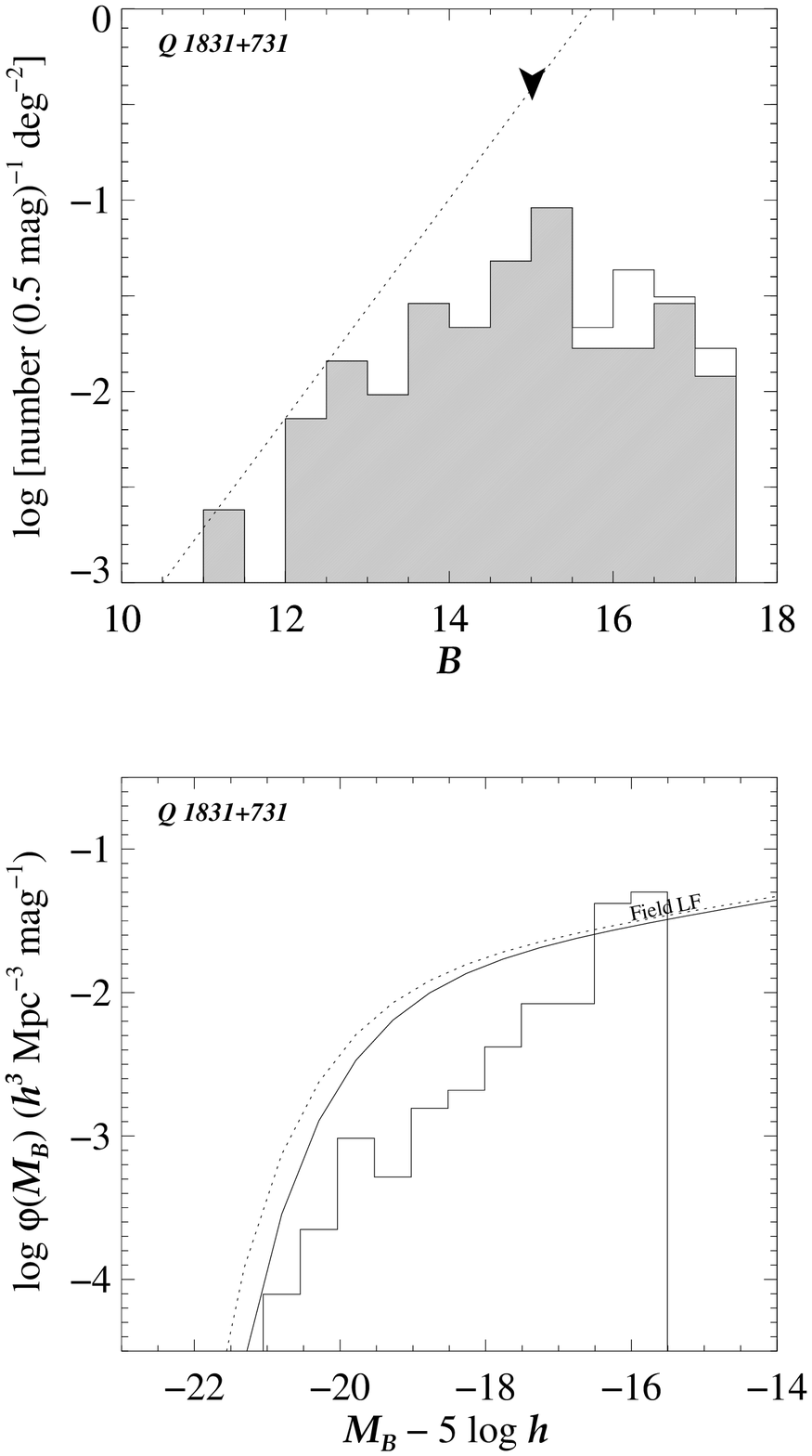,height=9cm,angle=0}\vspace*{-9cm}
\hspace*{4cm}{\psfig{figure=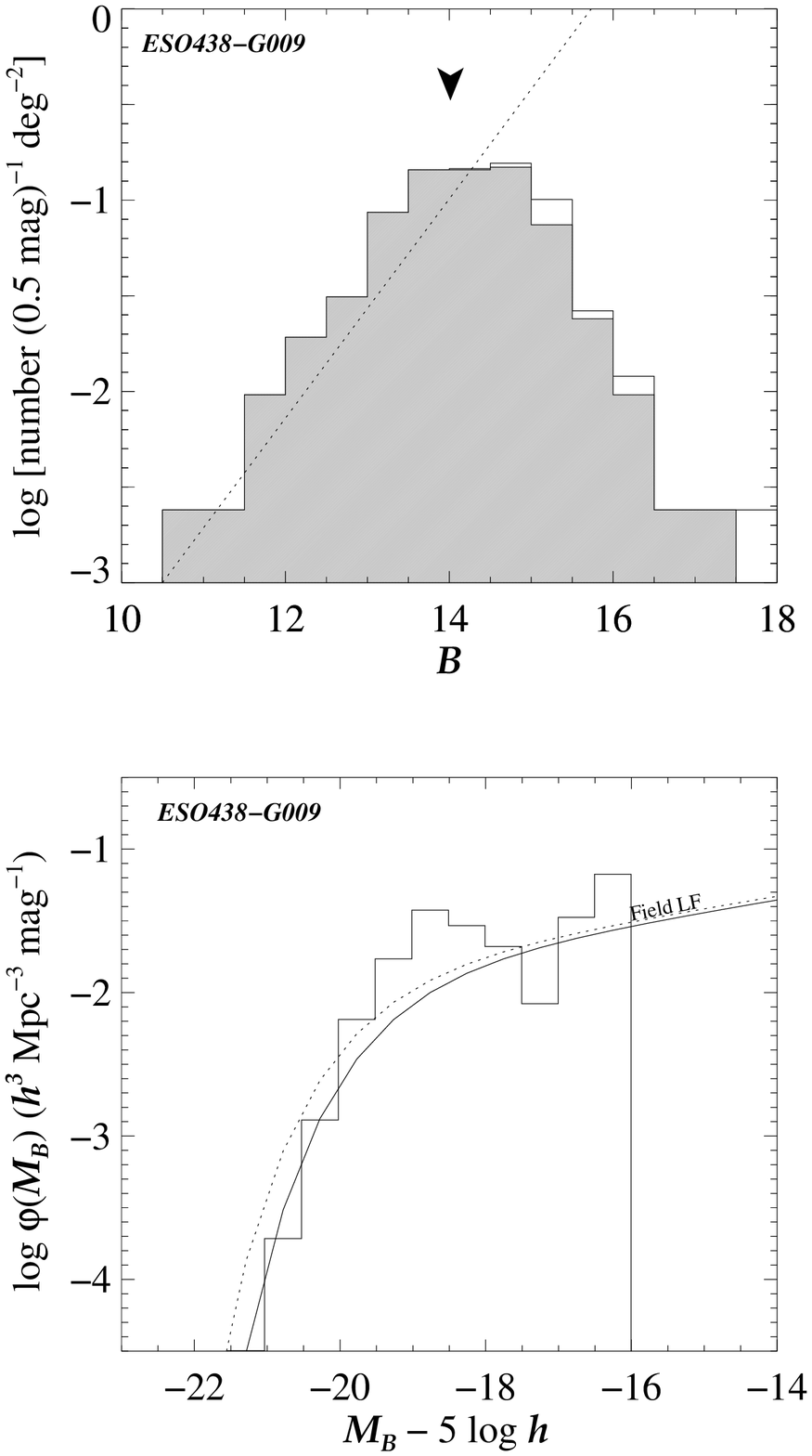,height=9cm,angle=0}}

\caption{\footnotesize Top: Galaxy number counts for all galaxies in a 691$'$ radius
($\equiv $2~\mpc\ at 1000~\kms ) for all objects cataloged in the
RC3. The shaded histogram represents the number of galaxies that also
have redshifts.  The galaxy number counts measured by \cite{yasuda01}
for galaxies in the SDSS data is shown with a dotted line. A
$\blacktriangledown$ marks the magnitude at which the RC3 seems to be
complete in this field.  Bottom: Derived differential galaxy
luminosity functions (LFs), compared to the $b_J$ field LF derived by
\cite{norberg01}. The lines shown are convereted to the $B$-band using 
$B = b_j +0.28(B-V)$ and assuming either $B-V = 0$ (solid line) or$ B-V
= 1$ (dotted line).
\label{fig_mags}}
\end{figure}

These magnitude distributions give us some confidence that we are complete
to some magnitude limit, $B_{\rm{lim}}$. We take that limit to be the
magnitude at which the number counts begin to turn over, or, more
precisely, the magnitude at the edge of the last bin where an increase in
magnitude is seen. These adopted limits are shown with arrows in
Figure~\ref{fig_mags} and are summarized in Table~\ref{tab_full}.

The distribution of $B$-band magnitudes includes galaxies of all redshifts
of course. A more precise measurement of the completeness of our galaxy
collation is the derivation of the galaxy luminosity function (LF), and its
comparison to the field LF. To estimate this, we have taken all the
galaxies brighter than the $B$-band limit  derived from the magnitude
distributions, and derived the classic differential LF

\begin{equation}
\phi(M)dm = \frac{ \sum_{i=1}^{N_g} N (M-dM/2\leq M_i \leq M+dM/2)}
{V_{\rm{max}}(M)}
\end{equation}

Here, $V_{\rm{max}}(M)$ is the volume corresponding to the maximum distance
a galaxy with magnitude $M$ which is bright enough to be included in the
sample could be observed \citep{schmidt68}. When used to derive the LF of a
sample of galaxies, this estimator is often
criticized for being over-sensitive to density fluctuations, giving poor
statistics when the objective is to derive the field LF
\citep[e.g.][]{will97}. In our case, however, this is an advantage, since
departures from the field LF (assuming our adopted magnitude limit is
reasonably accurate) can tell us directly how sparse or dense the galaxy
distribution is along a line of sight. 

LFs for our fields are shown in the lower panels of
Figure~\ref{fig_mags}. Over-plotted is the LF from the 2dF galaxy
redshift survey derived by \citet{norberg01}, which has the form of a
Schechter LF with $M^{*}=-19.67$, $\Phi^{*}=1.71\times10^{-2}$ and
$\alpha = -1.21$.  These values are derived in the $b_j$-band, but can
be converted to $B$ magnitudes using the relation $B = b_j +
0.28(B-V)$ \citep{blair82}. Since we have no colors for galaxies
selected from the RC3, we simply convert the 2dF $M^{*}$ value to a
$B$-band value by adding $0.3(B-V)$ and considering two cases, $(B-V) = 0$ and $(B-V)
= 1$.  When galaxies are counted in this way,
Figure~\ref{fig_mags} shows some interesting results. For Q1831+731,
there appears to be a genuine deficiency of galaxies for most
magnitudes compared to the field. (Note that this conclusion still
rests on whether the field has truly been sampled to the adopted
magnitude limit.) For MCG+10$-16-111$ the bright end of the LF closely
follows the field LF for bright galaxies until $M\sim-18$ when it
rises sharply, suggesting a preponderance of fainter galaxies, as is
seen in many clusters \citep[e.g.][]{yagi02}. The
rest of our fields, however,  match well the field LF.

\addtocounter{figure}{-1}

\begin{figure}
\hspace*{-1cm}\psfig{figure=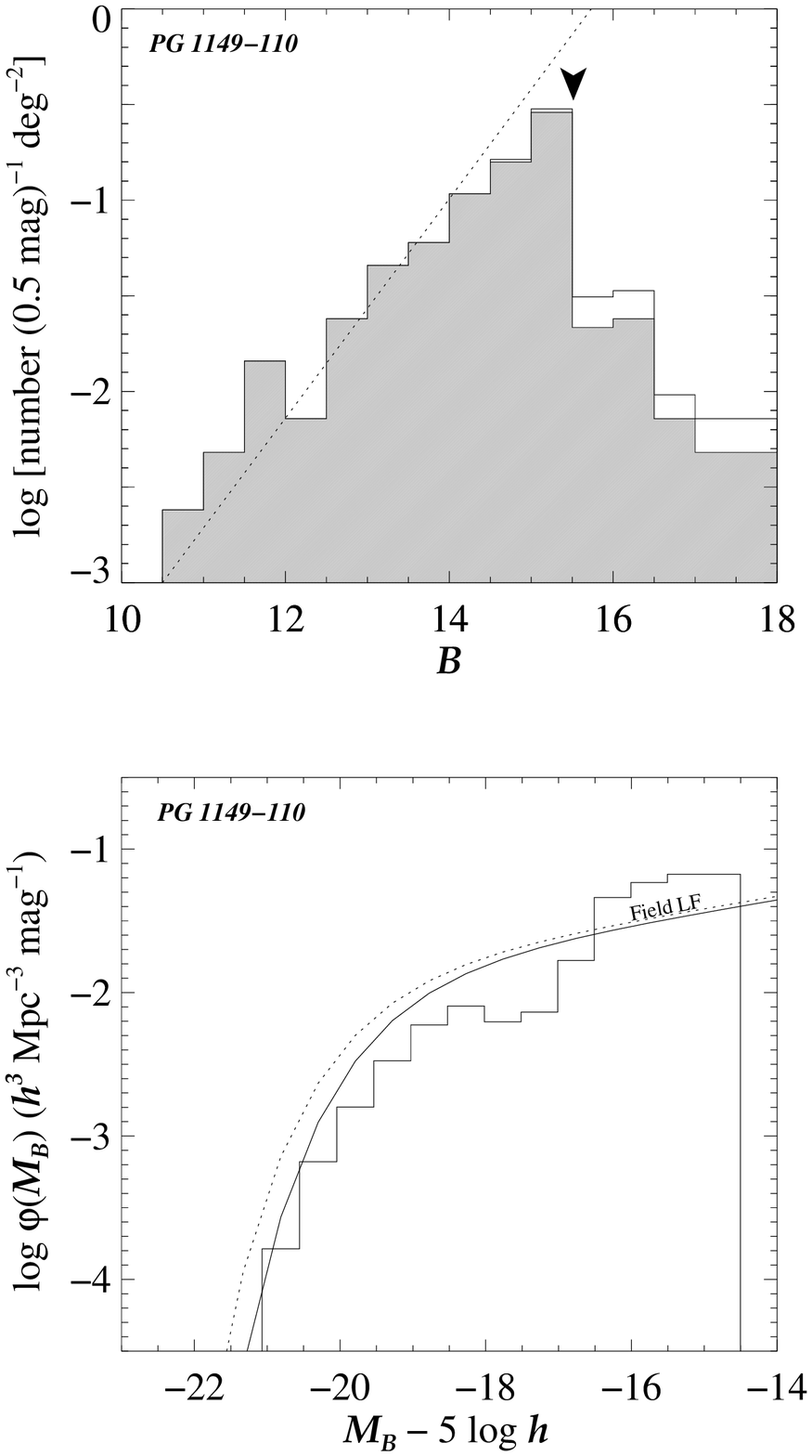,height=9cm,angle=0}\vspace*{-9cm}
\hspace*{4cm}{\psfig{figure=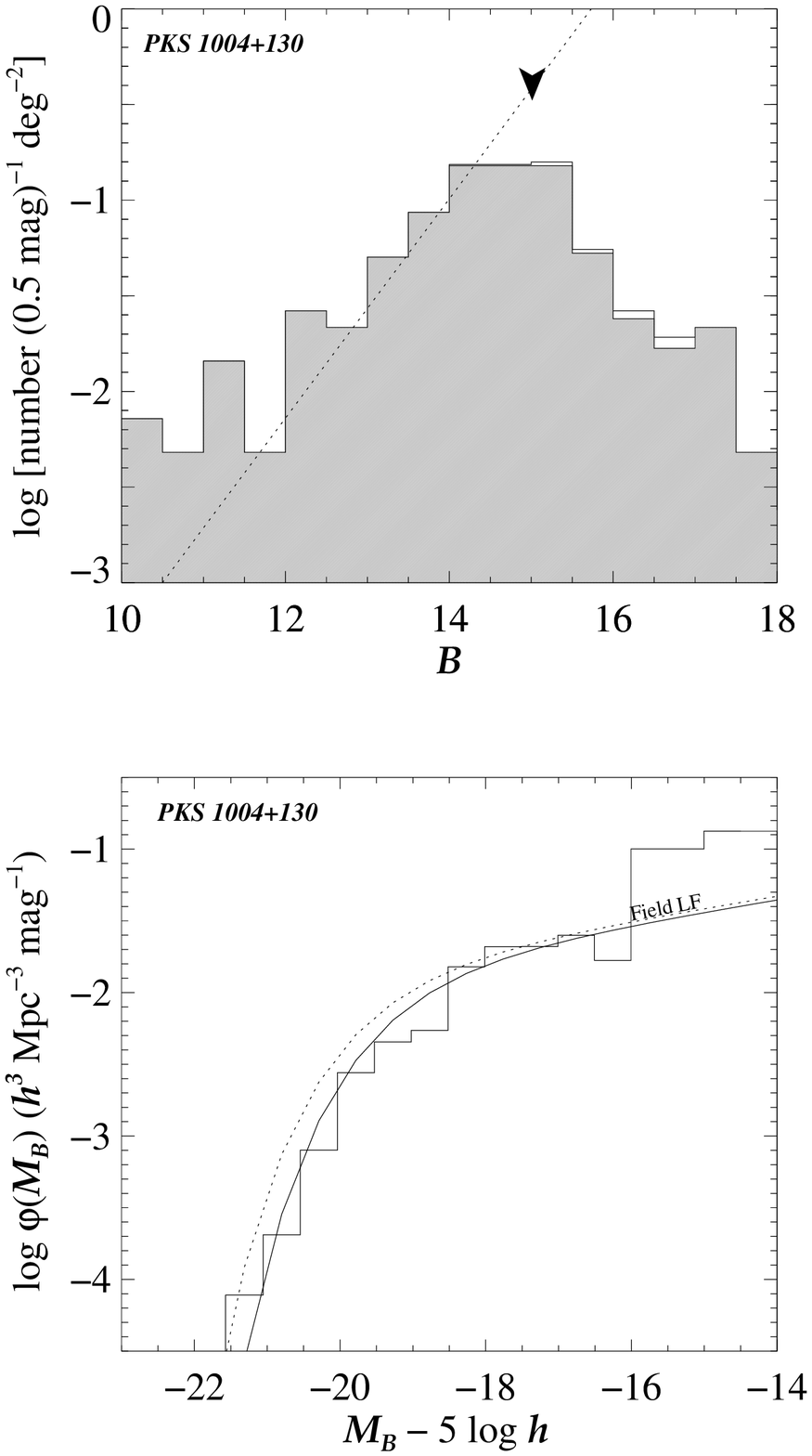,height=9cm,angle=0}}
\hspace*{-1cm}\psfig{figure=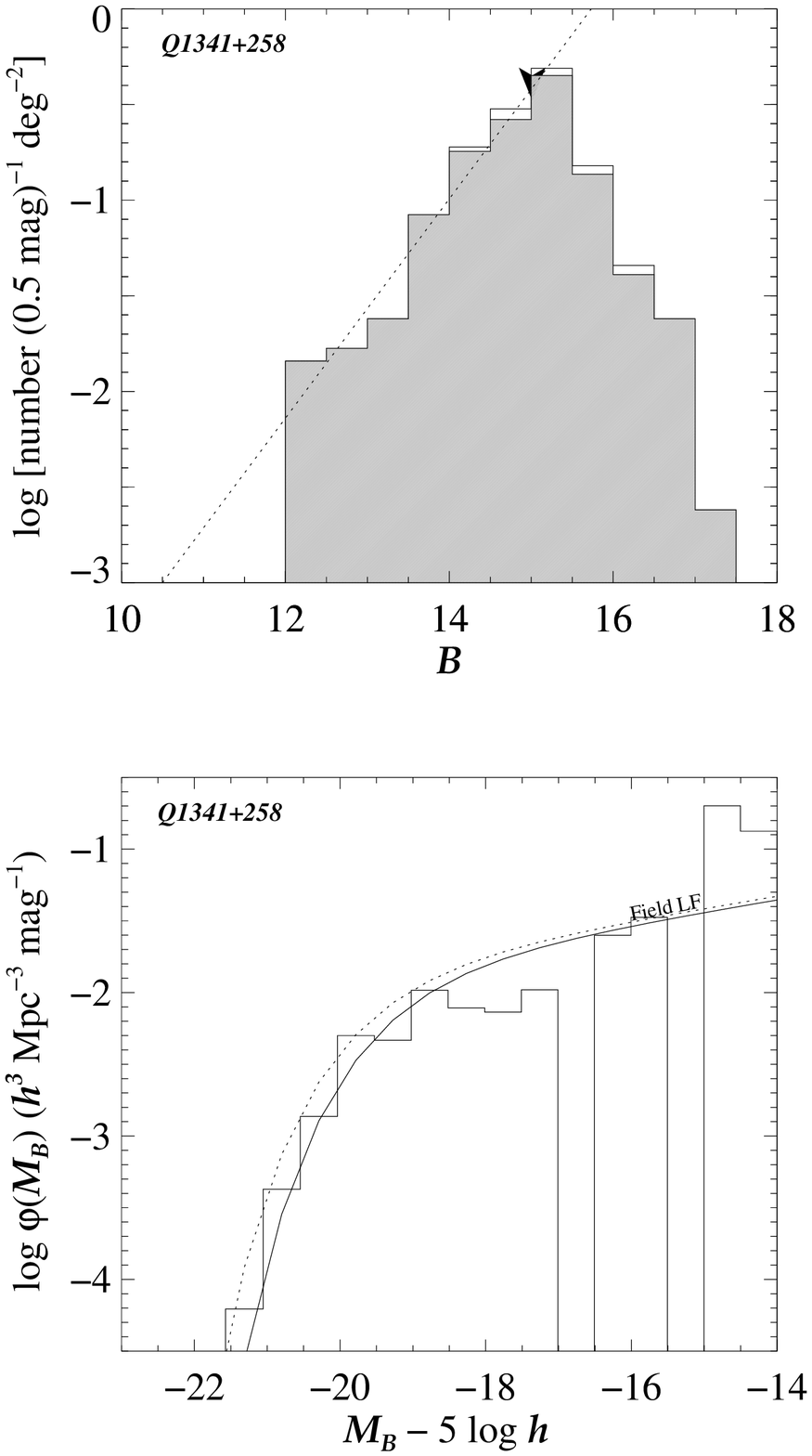,height=9cm,angle=0}
\caption{\footnotesize cont.}
\end{figure}

Our ultimate aim of course is to measure the density of galaxies
around a \lya\ system, to see if this correlates with \lya\
equivalent width or H~I column density. To derive $n$, the space
density of galaxies brighter than an absolute magnitude limit,
$M_{\rm{lim}}$, we simply calculate, for each field,

\begin{equation}
n  = \frac{ \sum_{N_g} N (M_B < M_{\rm{lim}},\rho <
\rho_{\rm{max}})}
{\pi \rho_{\rm{max}}^2 (d_{cz+500} - d_{cz-500})}
\end{equation}

i.e. we count the number of sample galaxies within a projected
separation of $\rho_{\rm{max}}$, which we take to be 2~\mpc\ (the
separation at which we are complete for $cz > 1000$~\kms ), and divide
by a volume defined by a cylinder of radius $\rho_{\rm{max}}$
stretching in length between a distance of $+500$~\kms\ and
$-500$~\kms\ from the \lya\ complex. ($d$ is the distance of a galaxy
at a redshift $z$ from us).  The volume defined in this way has a
length along the line of sight which is greater than the transverse
distance ($\rho_{\rm{max}}$). We would like the length along the line
of sight to be of order $\pm \rho_{\rm{max}}$ of the \lya\ system's
redshift, in order to be sure of sampling just the region of space
close to a galaxy.  However, with the peculiar velocity of a galaxy
likely to be a significant fraction of its systemic velocity, we
cannot take $cz_{\rm{gal}}$ as a precise measure of a galaxy's
distance from us. Distortions in redshift space due to the peculiar
velocities of galaxies in clusters are well known, and we must allow
for that distortion in counting galaxies along the line of sight.  We
adopt a dispersion of $\pm500$~\kms\ since this is about the maximum
observed for groups and poor clusters.  It also the same velocity
range (and hence volume) that we adopted in \S\ref{sect_fos} over
which we effectively summed equivalent widths and column densities of
individual \lya\ components, resulting in our estimates of $W_T$ and
$N_T$.  We obtain very similar results if we take a dispersion of only
$\pm300$~\kms , however. The ranges used are shown in
Table~\ref{tab_full}, and include exactly the same components used to
compare total equivalent widths with impact parameters in
\S\ref{sect_fos}.

$M_{\rm{lim}}$ is calculated from $B_{\rm{limit}}$ at a velocity
500~\kms\ greater than the \lya\ absorption system. The resulting
values are given in Table~\ref{tab_full}. Fortuitously, the range in
$M_{\rm{lim}}$ is small, and we can change $M_{\rm{lim}}$ to be a
fixed value so that we are counting galaxies all brighter than a
single value in each field. Table~\ref{tab_full} shows that using
$M_{\rm{lim}} = -17.5 $ is convenient, hence we calculate $n$ for
galaxies brighter than this. Note that since this limit is not
particularly faint, we sometime exclude in the census the galaxies we
originally selected to probe.

\begin{figure}
\hspace*{-1.5cm}\psfig{figure=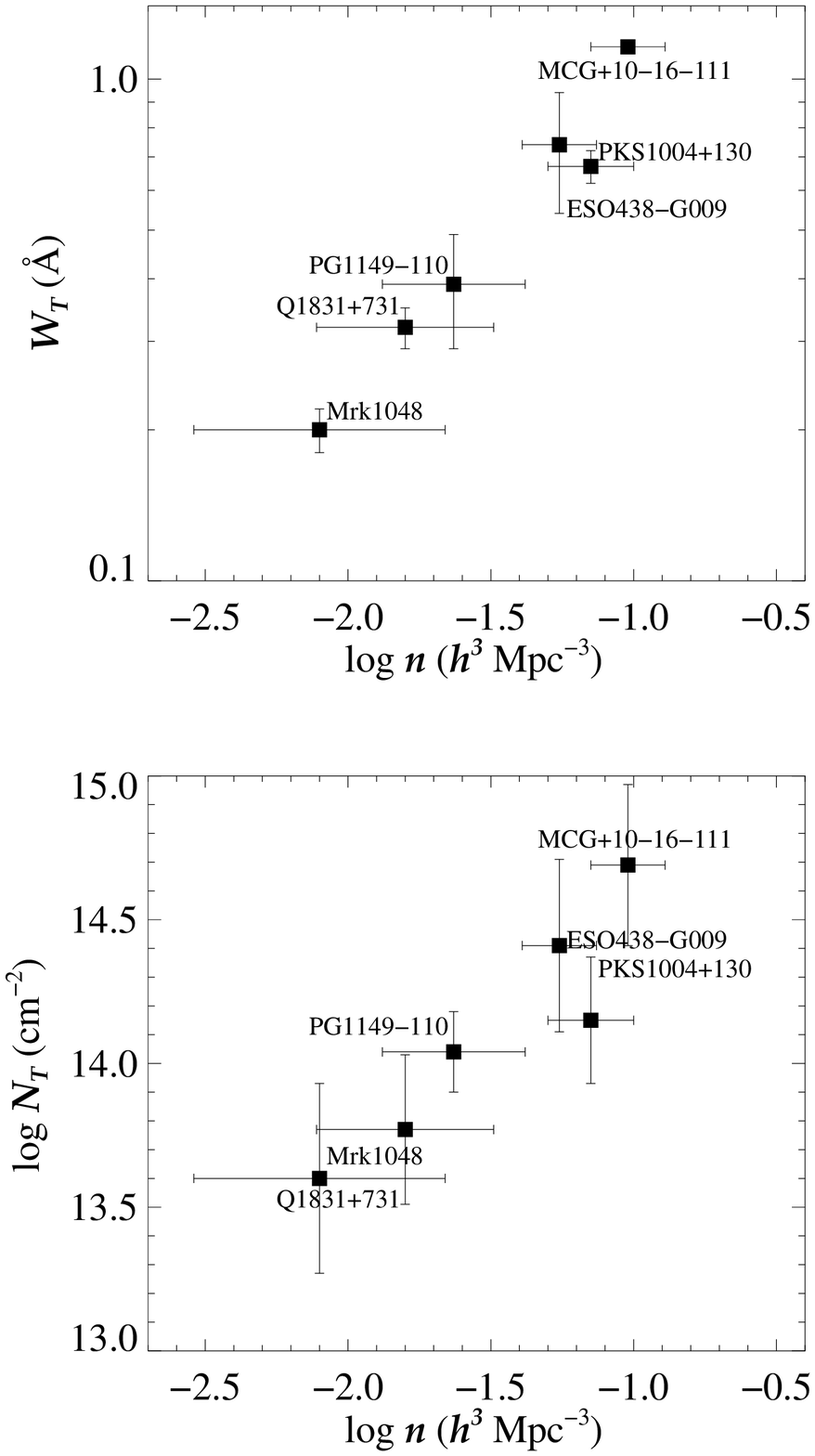,height=16cm}
\caption{\footnotesize Plot of the total equivalent width, $W_T$, 
(top) and total column density, $N_T$, (bottom) with 
the volume density of galaxies brighter than $M_B=-17.5$, $n$, in a cylinder of radius 
2~\mpc\ and length $\pm 500$~\kms\ from the center of a \lya\ complex.  The names of the background
probes are labelled. \label{fig_lss}}
\end{figure}

The top panel of Figure~\ref{fig_lss} shows the derived galaxy density
plotted against the integrated equivalent width, $W_T$. The errors in
$n$ are simply those from counting statistics, $\sqrt{n}$. The figures
shows that $W_T$ is correlated with $n$.  The bottom panel of
Figure~\ref{fig_lss} shows that the same correlation holds true if the
integrated column density, $N_T$, is also plotted against galaxy
density.  Hence, the higher the density of $M_B < -17.5$ galaxies in a
given volume, the stronger the equivalent width of \lya\ absorption
and H~I column density measured over 1000~\kms.

In the above analysis, we were interested in \lya\ complexes which
arose at the same velocities as galaxies close to the line of
sight. What about the remaining \lya\ lines in the STIS spectra? If we
are suggesting that the determining factor of the strength of a \lya\
complex is galaxy density, we should at least explore whether the same
relationship holds for \lya\ lines {\it not} selected because of the
proximity of a galaxy to the line of sight.

\begin{figure}
\hspace*{-1.5cm}\psfig{figure=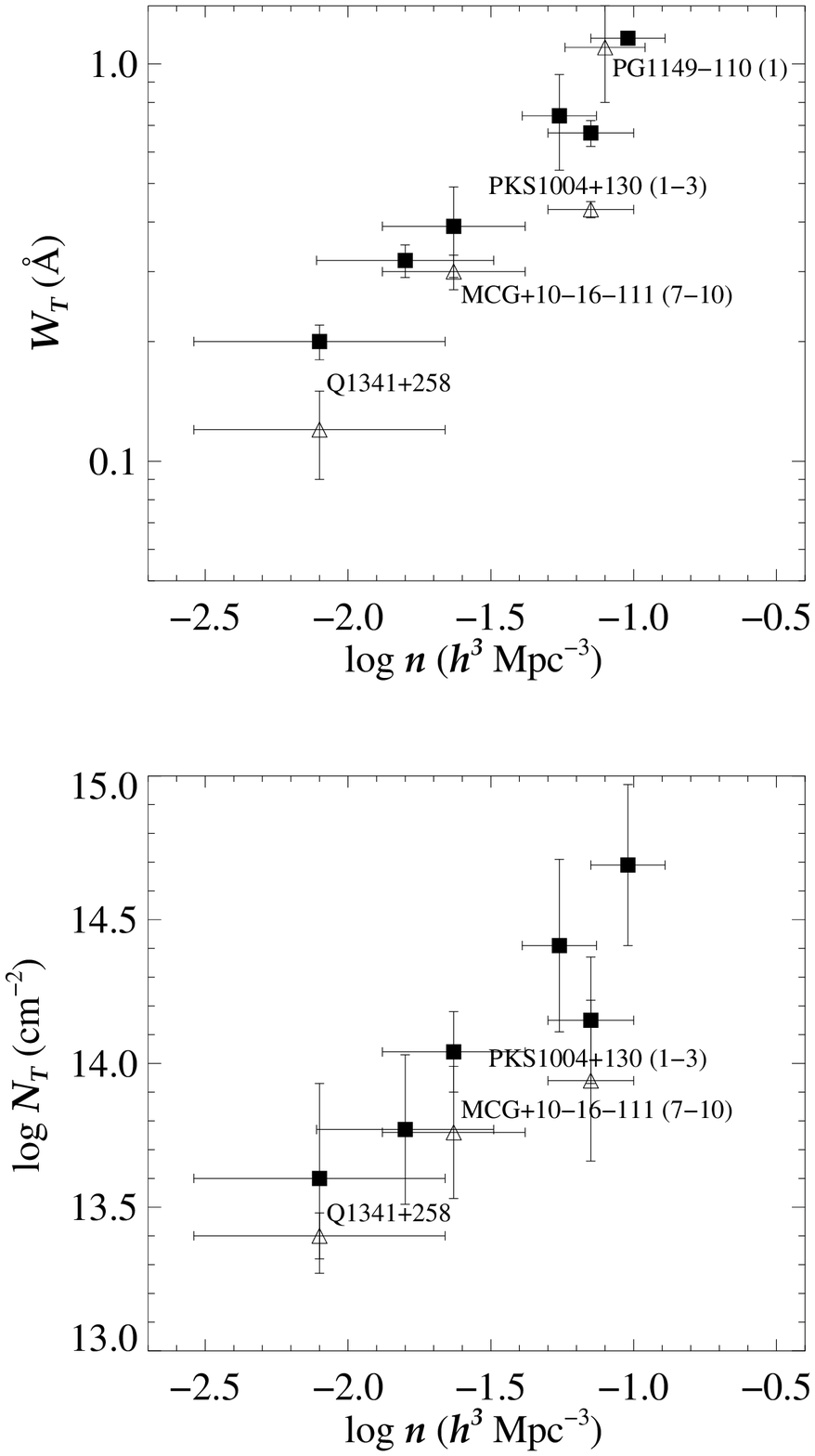,height=16cm}
\caption{\footnotesize Same as Figure~\ref{fig_lss} but now including all \lya\ complexes  detected in our
survey with redshifts $>1000$~\kms . Only the names of the four additional systems, shown as triangles,
 are labelled. No reliable 
column density is available for component 1 of PG~1149$-$110.
\label{fig_lss2}}
\end{figure}

We first impose the selection that all \lya\ lines must have redshifts
$>1000$~\kms , as we adopted above. The additional \lya\ lines are
then: components $1-3$ towards PKS~1004+130, $7-10$ towards
MCG~10$-$16$-$111, component 1 towards PG~1149-110, and the single
line we found towards PG~1341+258. Again, we need to sum these
components over a 1000~\kms\ range to replicate the procedures we
adopted above.  Choosing components to include is fairly
straightforward, and we sum equivalent widths and column densities of
components $1-3$ towards PKS~1004+130 and $7-10$ towards
MCG~10$-$16$-$111. We then repeat the calculation of identifying
galaxies within the volume defined above. Table~\ref{tab_full} shows
that we do not reach the $M_{\rm{lim}} = -17.5$ towards MCG~10$-$16$-$111
given $B_{\rm{lim}}$ for that field, and hence $n$ may be only a lower
limit. Conversely, we go too deep for the lower redshift \lya\ lines
towards the other three sightlines. In order to compare our results
with the densities shown in Figure~\ref{fig_lss}, we impose the limit
of $M_{\rm{lim}} = -17.5$ when counting galaxies. 
The results are shown in Figure~\ref{fig_lss2} where triangles
represent the four additional \lya\ line complexes, each labelled to show
which components were included. We also include the data from
Figure~\ref{fig_lss} as squares, but without labels, in order to avoid
over-crowding in the diagrams. From  Figure~\ref{fig_lss2} it can be
seen that the additional absorption complexes follow the correlation
we have found, within the uncertainties.

We conclude that there is suggestive evidence in favour of the density
of \lya\ components along a sightline being related to the volume
density of galaxies brighter than $-17.5$ within a few Mpc of a
sightline. Clearly, there are sufficient data in the HST Archive to
test this conclusion further using very low redshift \lya\ lines
detected in FOS, GHRS, and STIS data. We defer such a test for a later
paper. We stress again that these conclusions rest on the assumption
that the RC3 is capable of providing a magnitude limited sample along
a sightline to the limits given in Table~\ref{tab_full}. We note
though that a visual inspection of the field tends to support our
analysis of the catalog. That is, it is easy to see that certain fields
contain more bright galaxies than others. We do note, however, that
the correlations in Figs~\ref{fig_lss} and \ref{fig_lss2} are quite
sensitive to the numbers of galaxies counted, simply because the
actual numbers are small. A more detailed analysis, which might involve
counting galaxies to fainter magnitude limits, or simply selecting
more sightlines, will test our conclusions more rigorously.

                          \section{Conclusions and Summary \label{sect_conclusions}}

By selecting background QSOs whose sightlines pass close to low
redshift galaxies in the local universe, we have detected \lya\
absorption lines within a few hundred ~\kms\ of the systemic velocity
of a galaxy in all cases. We conclude that a line of sight which
passes within an impact parameter of $26-200$~\h\ of a galaxy is
likely to intercept low column density neutral hydrogen with $\log
N$(H~I)$\:\apg\: 13$. The ubiquity of detections implies a covering
factor of $\simeq 100$~\% around galaxies selected in this way. We
find that \lya\ absorption lines are actually composed of individual
components spread out in velocity over a range sometimes as large as
800$-$900~\kms .

Along the sightlines towards two probes, the high resolution and high
S/N of our STIS data reveal components which are unusually broad for
low-redshift \lya\ lines, with Doppler parameters $\sim 150$~\kms. If
these widths reflect the kinetic temperature of the absorbing gas,
then the gas must be at temperatures of $1-2\times10^{6}$~K.

In general, there are two possible explanations for the \lya\
absorption we see. The first is that absorbing gas may be directly
associated with the galaxies we chose to probe. Deposition of gas far
from stellar populations of a galaxy may be due to internal
interstellar processes within galaxies (e.g. superwinds from episodes
of intense star formation, or gas cycling between the disk and halo in
a galactic fountain) or perhaps from more external, dynamical
processes, such as galaxy-galaxy interactions, or galaxy-dwarf
accretion. The second explanation is that some or all of the absorbing
components are not physically located around the probed galaxy at
all. Instead, the absorbing gas simply traces the same large-scale
gravitational structures that the galaxies inhabit.

The amount of data presented in this paper is insufficient to
distinguish categorically between these two scenarios. However,
several of our results appear to indicate that the second of these
interpretations is at least as plausible as the first. 

Although it seems relatively straightforward to match an individual
absorption line with a given galaxy, we have discussed in
\S\ref{sect_corrs} how such a one-to-one assignment is made uncertain
by the multi-component structure of many of the \lya\ lines
detected. Furthermore, galaxies are often found in loose groups and
clusters. Although we targeted galaxies which appeared isolated, we
subsequently discovered that there is usually more than one galaxy
within our imposed impact parameter of 200~\h\ from the QSO sightline.
By degrading our STIS data to the resolution of the FOS, the
multicomponent nature of our absorption systems is largely removed. By
matching these degraded lines (effectively summing individual
component equivalent widths and column densities over 1000~\kms\ to
produce an equivalent width $W_T$ and column density $N_T$) to the
galaxy closest to the sightline, and ignoring any companions, we
are able to reproduce the anti-correlation of \lya\ equivalent width
and impact parameter found at higher redshift. This suggests that the
population of absorbing galaxies studied at high-$z$ is no different
than the nearby galaxies we selected to probe.  

We find that our anti-correlation is weakened by correcting for galaxy
luminosity in the manner prescribed for the higher redshift systems,
$W\propto L^{0.4}$, yet appears to improve for $W\propto
L^{-0.3}$. Such an relationship is surprising. It is possible to
envisage a scenario whereby brighter galaxies have used up all the
available neutral hydrogen in their immediate vicinity, resulting in
brighter galaxies being associated with weaker \lya\ lines at a
particular radius. Nevertheless, such a behaviour is significantly at
odds with the results at higher redshift, and would imply that our
population of absorbing galaxies had little to do with those found at
higher-$z$. It also seems largely incompatible with the result
discussed below and \S\ref{sect_lss}, that $W_T$ correlates with the
volume density of moderately bright galaxies.  We note that the
improved correlation becomes significant largely due to the correction
for a single galaxy, UGCA~226 in the field of ESO~438-G009 (compare
the top and bottom panels of Figure~\ref{fig_corrs1}). Also, the
rank correlation tests suggest the correlation is only marginally
significant, at somewhere between the 2.2$\sigma$ (Spearman) and
2.8$\sigma$ (Kendall) level. We suspect therefore that the improvement
in the correlation after correcting for galaxy luminosity is
coincidental, a result of small number statistics, but we cannot rule
out the relationship is real. Obvioulsy, further data are required to
explore this issue in more detail.

Collating all available galaxy redshifts along each line of sight has
shown that many of our selected galaxies are actually members of
moderately rich groups, and we could equally well conclude that low
redshift \lya\ lines are associated with groups of galaxies. We have
investigated this alternative interpretation quantitatively, by
defining a simple measure of the volume density of galaxies, $n$, evaluated
at the same distance as a given galaxy/\lya-complex. Assuming that the
RC3 provides a magnitude limited sample (to the depths determined in
Table~\ref{tab_full}),  a correlation appears to exist between
$n$ and both the total \lya\ equivalent width and H~I column density
of an absorption complex (Figures~\ref{fig_lss} and \ref{fig_lss2}).
This is highly suggestive evidence for an association of the \lya\
absorption lines with larger-scale structures rather than with individual
galaxies, as predicted by hydrodynamical simulations of the
growth of structure in currently favoured cosmologies. The evidence is
strengthened by the fact that the same correlation seems to apply
irrespective of whether there is a bright galaxy within 200~\h\ of a
sightline.

This second scenario has a parallel at high redshift, where
\citet{Adel02} also see a strong correlation between the optical depth
in the \lya\ forest and the overdensity of galaxies.  From their
observations, they find that the forest is actually cleared out in the
immediate vicinity of galaxies (at impact parameters of 100--200~\h )
which appears to be different from the galaxies studied in our sample
(presumably from outflows generated by active star-forming galaxies at
that epoch).  However, on the same scales that we have been
considering, 1--2~\mpc , there is an excess of \lya\ absorption
closely related to the density of galaxies in the high redshift
sample. The parallel is intriguing because the high-$z$ and low-$z$
\lya\ forest have been considered to be different populations, given
their markedly different redshift evolutions \citep[e.g.][]{KP14}.
Possibly, the two populations are closely related, as has been
suggested by results from the hydrodynamical simulations of
\citet{dave99}.

Can our conclusions be reconciled with other results from studies of
$z<1$ \lya\ lines?  One corollary would be that the strongest lines
seen in FOS spectra should be associated with the highest density regions
of galaxies, in rich groups or clusters.  (Of course, at sufficiently
small impact parameters we would expect the outer regions of
individual galaxy disks to be intercepted directly.)  It should be
possible to test this prediction by completing the redshift surveys which have
been underway for several years, obtaining redshifts of {\it all}
galaxies within a reasonably large radius (e.g. 2~\mpc\ $\equiv 11$
arcmin at $z=0.3$) down to a specified magnitude limit. Unfortunately,
the amount of telescope time required to fully sample a field may be
hard to come-by. A field would need to be sampled to a depth where the
galaxy luminosity function was adequately sampled at the absorber
redshift.  For example, if the field luminosity function used in
\S\ref{sect_lss} is valid at higher redshift, there would be only
half a dozen or so $M=-18$ galaxies in the cylindrical volume we used
to measure galaxy volume density. Unfortunately, an $M=-18$ galaxy
would have an observed magnitude of $m\sim22$ at $z=0.3$, meaning that
several hundred galaxies with magnitudes brighter than this limit
(which would include foreground and background galaxies) would have to
be surveyed to properly sample the field at the required redshift.

The anti-correlation of equivalent width and impact parameter found at
higher redshift might also be understood in terms of galaxy density:
stronger lines would arise in regions of higher galaxy density, where
the probability of finding a galaxy close to the sightline would be
high because of the increase in the number of galaxies per unit
area. It is likely that this effect has already been seen in the
simulations of, e.g., \citet{dave99}.  If galaxy density principally
governs \lya\ equivalent width, then why should \lya\ equivalent width
correlate with the luminosity of the galaxy closest to the QSO line of
sight \citep[Figure 2 of][]{chen01}?  One possibility is that galaxies
in high density regions (giving rise to strong \lya\ lines) may be
brighter than those in low density regions (which produce only weak
\lya\ lines). So QSO sightlines which intercept richer groups of
galaxies are more likely to find brighter galaxies, which are
themselves likely to be closer to a sightline due to the higher
surface density of galaxies.  The exact shape of the luminosity
function in groups and clusters, and its variation between
environments of different densities, is still unclear, but there is
evidence for an increase in the luminosity of galaxies in denser
environments \citep[e.g.][]{garilli99, yagi02, norberg02}.

Whether these effects are significant enough to recreate the
correlations seem for the high redshift galaxy sample remains uncertain.
Only more comprehensive redshift surveys around high-$z$ QSO
fields, and more detailed studies of the absorbing population of
galaxies in the nearby universe, might eventually help us to understand
whether galaxy density and environment really are important factors
in governing the existence and conditions of \lya\ clouds.

\acknowledgements

We'd like to thank Todd Tripp, Ed Jenkins and Brenda Frye for helpful discussions,
and Norman Grogin for providing us with important UZC galaxy spectra.
Thanks also to Joss Bland-Hawthorn for providing Perspective,
and especially to the referee for a careful and insightful reading of the
paper. Support for this work was provided through grant GO-08316.01
from the Space Telescope Science Institute (STScI), which is operated
by the Association of Universities for Research in Astronomy, Inc.,
under NASA contract NAS5-26555. The NASA/IPAC Extragalactic Database
(NED) is operated by the Jet Propulsion Laboratory, California
Institute of Technology, under contract with the NASA. The Digitized
Sky Surveys were produced at the STScI under U.S. Government grant
NAGW-2166.

\clearpage

\bibliographystyle{apj}
\bibliography{apj-jour,bib1}

\end{document}